\documentclass[submission,nomarks]{dmtcs-episciences}

\usepackage[utf8]{inputenc}
\usepackage{subfigure}

\usepackage[round]{natbib}

\usepackage{makecell}
\usepackage{amsmath}
\usepackage{amssymb}
\usepackage{aliascnt}
\usepackage{tikz}
\usepackage{refcount}
\usepackage{cleveref}

\usepackage{array}
\usepackage{tabularx}
\usepackage{makecell}
\usepackage{threeparttable}

\newcolumntype{Y}{>{\centering\arraybackslash}X}

\newcommand{\dist}{\operatorname{dist}}

\newcommand{\pol}{\operatorname{Pol}}
\newcommand{\ppol}{\operatorname{pPol}}
\newcommand{\rcsp}{\mathrm{RCSP}}
\newcommand{\rcspc}{\mathrm{RCSP}_{\mathrm C}}

\newcommand{\false}{\mathbf{f}}
\newcommand{\bs}{\boldsymbol}

\newcommand{\maj}{\operatorname{maj}}

\newcommand{\Mal}{M}
\newcommand{\dom}{\operatorname{dom}}
\newcommand{\csp}{\mathrm{CSP}}
\newcommand{\cspc}{\mathrm{CSP}_{\mathrm C}}
\newcommand{\pto}{\rightharpoonup}
\newcommand{\inv}{\operatorname{Inv}}

\newcommand{\scb}{\mathrm{SCB}}
\newcommand{\sor}{\mathrm{SOF}}
\newcommand{\snand}{\mathrm{SNF}}
\newcommand{\var}{\mathrm{VAR}}
\newcommand{\assi}{t}

\newif\ifconference
\conferencefalse

\makeatletter
\def\iddots{\mathinner{\mkern1mu\raise\p@
		\hbox{.}\mkern2mu\raise4\p@\hbox{.}\mkern2mu
		\raise7\p@\vbox{\kern7\p@\hbox{.}}\mkern1mu}}
\makeatother

\newtheorem{theorem}{Theorem}[section]
\newaliascnt{lemma}{theorem}
\newtheorem{lemma}[lemma]{Lemma}
\aliascntresetthe{lemma}

\newaliascnt{proposition}{theorem}
\newtheorem{proposition}[proposition]{Proposition}
\aliascntresetthe{proposition}

\newaliascnt{corollary}{theorem}
\newtheorem{corollary}[corollary]{Corollary}
\aliascntresetthe{corollary}

\newaliascnt{definition}{theorem}
\newtheorem{definition}[definition]{Definition}
\aliascntresetthe{definition}

\newaliascnt{example}{theorem}
\newtheorem{example}[example]{Example}
\aliascntresetthe{example}

\newaliascnt{remark}{theorem}
\newtheorem{remark}[remark]{Remark}
\aliascntresetthe{remark}

\newaliascnt{claim}{theorem}

\aliascntresetthe{claim}

\crefname{theorem}{Theorem}{Theorems}
\Crefname{theorem}{Theorem}{Theorems}

\crefname{lemma}{Lemma}{Lemmas}
\Crefname{lemma}{Lemma}{Lemmas}

\crefname{proposition}{Proposition}{Propositions}
\Crefname{proposition}{Proposition}{Propositions}

\crefname{corollary}{Corollary}{Corollaries}
\Crefname{corollary}{Corollary}{Corollaries}

\crefname{definition}{Definition}{Definitions}
\Crefname{definition}{Definition}{Definitions}

\crefname{example}{Example}{Examples}
\Crefname{example}{Example}{Examples}

\crefname{remark}{Remark}{Remarks}
\Crefname{remark}{Remark}{Remarks}

\crefname{claim}{Claim}{Claims}
\Crefname{claim}{Claim}{Claims}

\crefname{section}{Section}{Sections}
\Crefname{section}{Section}{Sections}
\crefname{subsection}{Section}{Sections}
\Crefname{subsection}{Section}{Sections}
\crefname{equation}{Equation}{Equations}
\Crefname{equation}{Equation}{Equations}
\crefname{algorithm}{Algorithm}{Algorithms}
\Crefname{algorithm}{Algorithm}{Algorithms}
\crefname{appendix}{Appendix}{Appendices}
\Crefname{appendix}{Appendix}{Appendices}

\title{Towards an algebraic approach to the reconfiguration CSP}
\author[Kei Kimura]{Kei Kimura%
\thanks{This work was partially supported by JSPS KAKENHI
Grant Number JP21K17700 and JST ERATO Grant Number JPMJER2301,
Japan.}}

\affiliation{
  Kyushu University, Japan}

\begin{document}
	\maketitle
	
	\begin{abstract}
		This paper investigates the reconfiguration variant of the Constraint Satisfaction Problem (CSP), referred to as the Reconfiguration CSP (RCSP). Given a CSP instance and two of its solutions, RCSP asks whether one solution can be transformed into the other via a sequence of intermediate solutions, each differing by the assignment of a single variable. RCSP has attracted growing interest in theoretical computer science, and when the variable domain is Boolean, the computational complexity of RCSP exhibits a dichotomy depending on the allowed constraint types. A notable special case is the reconfiguration of graph homomorphisms---also known as graph recoloring---which has been studied using topological methods.
		We propose a novel algebraic approach to RCSP, inspired by techniques used in classical CSP complexity analysis. Unlike traditional methods based on total operations, our framework employs partial operations to capture a reduction involving equality constraints. This perspective facilitates the extension of complexity results from Boolean domains to more general settings, demonstrating the versatility of partial operations in identifying tractable RCSP instances. 
	\end{abstract}
	\section{Introduction}\label{sec:introduction}
	\ifconference
	\else
	The constraint satisfaction problem (CSP) is a problem of determining whether there exists a solution, which is an assignment of values from a given domain to variables that satisfies all the given constraints.
	Due to its modeling capabilities and the ability to describe a variety of problems, 
	the CSP has been studied in a wide range of fields such as mathematics, artificial intelligence, and computer science~\citep{Dec03,BKW17}.
	Various variants of the CSP have been studied, including approximation~\citep{KSTW01,Raghavendra08}, optimization~\citep{Ziv12}, counting~\citep{DR10}, promise~\citep{BG21,BBKO21}, and reconfiguration~\citep{GKMP09}.
	
	In the analysis of computational complexity of the CSP, 
	the constraint language restriction-based approach has been actively studied.
	That is, the study analyzes the computational complexity of the CSP when restricting the types of constraints that can be allowed in an instance.
	This is because the restriction of the constraint language allows various problems to be described individually. 
	For example, the Boolean satisfiability problem (SAT), itself an important research subject in computer science, and the graph coloring problem, itself an important research subject in graph theory and computer science, can be expressed as problems of this form.
	Algebraic approaches have been most successful in analyzing the computational complexity of the CSP based on constraint language restrictions.
	Broadly speaking, in the algebraic approach, the computational complexity of the CSP is characterized by the operations under which the constraint language is invariant.
	In fact, a computational complexity dichotomy theorem for the CSP based on constraint language restrictions has been shown using an algebraic method~\citep{Bul17,Zhu17,Zhu20}.
    Algebraic methods have also been employed in the study of optimization versions of the CSP~\citep{Ziv12}, the promise CSP~\citep{BG21,BBKO21}, and the fine-grained computational complexity of the CSP~\citep{JLNZ17,LW21}.
    
	In this study, we focus on the reconfiguration CSP (RCSP).
	The RCSP is, given an instance of the CSP and two solutions of the instance, 
	to determine if there exists a sequence of steps such that each step
	produces an intermediate solution.
	Problems of this kind have been actively studied in the fields of theoretical computer science, algorithms, and graph theory as \emph{combinatorial reconfiguration} in recent years~\citep{IDHPSUU11}.
	Since the RCSP is about the connectivity of the solution space, and since the connectivity of the solution space, especially in the Boolean case, is related to the performance of algorithms for SAT such as Walk SAT and DPLL, its analysis has been done especially for random instances, and a worst-case analysis is performed by \citet{GKMP09} and \citet{Sch14}.
	Furthermore, the graph homomorphism problem, which generalizes the graph coloring problem, is a special case of the CSP. 
	Its reconfiguration variant, known as graph recoloring, has been the subject of active research, particularly in analyzing its computational complexity when the codomain graph is fixed~\citep{CvJ11,BC09,BMMN16}---corresponding to the restriction of the constraint language in the CSP framework.
	The RCSP provides a unified framework for addressing these problems.
	
	For RCSPs, a systematic method for analyzing computational complexity under restricted constraint languages remains unknown. 
	In particular, to the best of the author's knowledge, no algebraic approach---such as those that have proven successful in the case of CSPs---has yet been applied. 
	Nevertheless, it is noteworthy that approaches based on topological methods have yielded meaningful results for the graph recoloring problem~\citep{BLS18,Wro20,LNS23,LMS25}.
	\fi 
	
	\paragraph*{Our contribution}
	\ifconference
	\else
	In this study, we introduce an algebraic approach, specifically an analysis using \emph{partial operations}, which are operations that are not defined for some inputs, into the computational complexity analysis of the RCSP based on the constraint language restriction.
	More precisely, we focus on partial operations under which a constraint language is invariant. 
	Here, the constraint language being invariant under a partial operation intuitively means that when applying the partial operation to several solutions, and the result of the operation is defined, then the result is also a solution in any instance of the CSP using the constraint language.
	
	First, we show that partial operations capture a reduction between RCSPs, assuming that the corresponding constraint languages allow the use of the equality constraint. Our result is motivated by the well-known fact in the CSP literature that partial operations characterize quantifier-free primitive positive definability, a characterization established by \citet{Gei68} and \citet{Rom81}. As a consequence, the computational complexity of RCSPs with equality is determined by the set of partial operations under which these are invariant.
	
	Next, we analyze the classes of RCSPs that can be solved in polynomial time in the Boolean case using partial operations.
	\citet{GKMP09} and \citet{Sch14} show that if a constraint language satisfies at least one of the following three conditions, then the RCSP can be solved in polynomial time: safely OR-free, safely NAND-free, and safely componentwise bijunctive.
	Here, safe OR-freeness and safe NAND-freeness are dual to each other (more precisely, one can be obtained from the other by swapping the two values in the Boolean domain) and can be treated equivalently in the analysis of computational complexity.
	First, we define a new partial operation, called the ordered partial Maltsev operation, derived from the well-known Maltsev operation using a total order on the Boolean domain. We show that safe OR-freeness is characterized by this operation. Furthermore, by extending it to larger domains, we obtain a new class of RCSPs that can be solved in polynomial time over arbitrary finite domains. Moreover, we show that this class encompasses various tractable classes studied in the CSP and related literature.
	Next, we show that the safe componentwise bijunctivity is also characterized by partial operations.
	However, we also show that this property cannot be characterized by any finite set of partial operations, in stark contrast to the case of safe OR-freeness.
	\fi 
	
	\ifconference
	\else
	\paragraph*{Related work}
	A seminal work of \citet{GKMP09} reveals that 
	the computational complexity of the Boolean RCSP with constants\footnote{\citet{GKMP09} referred to the RCSP as the $st$-connectivity problem.} exhibits the following dichotomy: 
	the problem is solvable in polynomial time if the constraints satisfy a certain property and otherwise PSPACE-complete, although the criterion for dividing the complexity is later revised by \citet{Sch14}.
	\citet{BBD+21} provide a sufficient condition for the solution space of CSP to be connected.
	\citet{HIZ18} conduct a detailed analysis of the computational complexity of RCSPs, particularly from the perspective of fixed parameter tractability, based on an approach that restricts the graphical structure of how constraints and variables interact, which is distinct from the approach of restricting the constraint language.
	
	The RCSP with a constraint language consisting of a single binary relation corresponds to the digraph recoloring problem. The computational complexity of $H$-recoloring has been extensively studied for various digraphs $H$.
	\citet{CvJ11} show that the $K_3$-recoloring, which is the reconfiguration version of the 3-coloring problem, is solvable in polynomial time.
	On the other hand, the $K_k$-recoloring is shown to be PSPACE-complete for each $k \ge 4$ by \citet{BC09}, exhibiting a complexity dichotomy theorem for the reconfiguration version of the graph coloring problem.
	Extending this dichotomy theorem, \citet{BMMN16} established a dichotomy theorem for the reconfiguration version of circular coloring.
	This problem is equivalent to the $C_{p,q}$-recoloring problem for the circular clique $C_{p,q}$, where $C_{p,q}$ is the graph with vertex set $\{0,1,\dots, p-1\}$ and edge set $\{ ij \mid q \le |i-j| \le p-q \}$.
	More specifically, \citet{BMMN16} showed that $C_{p,q}$-recoloring is solvable in polynomial time when $2 \le p/q < 4$, and is PSPACE-complete when $p/q \ge 4$.
	On the hardness side of computational complexity, it has also been shown that $H$-recoloring is PSPACE-complete when $H$ is a $K_{2,3}$-free quadrangulation of the 2-sphere that is not the 4-cycle $C_4$~\citep{LNS20}
	
	As a major result concerning the polynomial-time solvability of $H$-recoloring for undirected graphs, \citet{Wro20} showed that $H$-recoloring is solvable in polynomial time when $H$ is square-free, that is, when it does not contain the 4-cycle $C_4$ as a subgraph. 
	Wrochna’s approach treats the solution space as a topological space and employs topological conditions that the reconfiguration sequence of colorings must satisfy, resulting in an algorithm that is both elegant and conceptually intriguing.
	Subsequent research has extended Wrochna’s topological approach, demonstrating that $H$-recoloring is solvable in polynomial time when $H$ satisfies any of the following conditions: 
	(i) $H$ is a reflexive digraph cycle that does not contain a 4-cycle of algebraic girth 0~\citep{BLS18}, 
	(ii) $H$ is reflexive and has girth at least 5~\citep{LNS23}, 
	(iii) $H$ is a loopless digraph that contains no 4-cycle of algebraic girth 0~\citep{LMS25}, and 
	(iv) $H$ is a reflexive digraph that contains neither a triangle of algebraic girth 1 nor a 4-cycle of algebraic girth 0~\citep{LMS25}.
	A comparison between these results and our results will be conducted in \cref{sec:SOR}.
	
	The use of partial polymorphisms derived from a total order on the domain has also been explored in the context of CSPs with unbounded domains and few variables, where \citet{JLdW26} uses such order-based partial polymorphisms to characterize tractable classes of constraint languages over infinite domains. Our ordered partial Maltsev operation can be viewed as another instance of this general principle, namely, extracting algorithmically useful structure from a total order on the domain via partial operations.
	
	
	\fi 
	
	\paragraph*{Organization}
	The structure of this paper is as follows. \cref{sec:preliminaries} introduces the formal definition of RCSPs and reviews previously known results. 
	\cref{sec:SOR} focuses on safe OR-freeness, while \cref{sec:SCB} discusses safe componentwise bijunctivity. 
	Finally, \cref{sec:conclusion} concludes the paper.
	
	\section{Preliminaries}
	\label{sec:preliminaries}
	
	In this section, we first define the Reconfiguration CSP (RCSP). Then, in \Cref{subsec:ppol}, we introduce partial operations and explain the reduction relationships between RCSPs. Furthermore, in \Cref{subsec:results-for-Boolean-RCSP}, we summarize known results in the context of Boolean RCSPs.
	
	\subsection{Definitions}
	For a positive integer $r$ we denote $\{1, \dots, r \}$ by $[r]$.
	Throughout the paper, $D$ denotes a finite domain whose size is equal or greater than two.
	
	\begin{definition}
		An \emph{$r$-ary relation on $D$} is a subset of $D^r$, where $r \ge 1$.
	\end{definition}
	
	\begin{definition}
		The following are some standard relations that will be used throughout the paper.
		\begin{itemize}
			\item The \emph{diagonal (or equality) relation} $\Delta_D$ on $D$ is a binary relation $\Delta_D = \{ (d,d) \in D^2 \mid d \in D \} \subseteq D^2$.
			\item The \emph{inequality relation} $\neq_D$ on $D$ is a binary relation $\neq_D = \{ (c,d) \in D^2 \mid c \neq d \} \subseteq D^2$.
			\item For each $d \in D$, let $C_{d}$ denote the singleton unary relation $C_d = \{ d \} \subseteq D$.
			\item The $r$-ary \emph{empty relation} $\emptyset^{(r)}$ is the empty set $\emptyset \subseteq D^r$, where $r \ge 1$.
		\end{itemize}
	\end{definition}
	
	\begin{definition}
		A \emph{constraint language} is a finite set of non-empty relations on $D$.
	\end{definition}
	
	\begin{definition}\label{def:varGamma-formula}
		\ifconference
		\else
		For a (not necessarily finite) set of relations $\varGamma$ on $D$, 
		a \emph{$\varGamma$-formula} 
		$I$ is a conjunction of the form
		\begin{align}\label{eq:CSP-instance}
			I = \bigwedge_{i = 1}^{m} R_i(x^i_1,\dots,x^i_{r_i}),
		\end{align}
		where $m$ is a positive integer, each $R_i$ is an $r_i$-ary relation from $\varGamma$, 
		and the $x^i_{j}$ are (not necessarily distinct) variables.
		\fi 
		
		For a (not necessarily finite) set of relations $\varGamma$ on $D$, 
		a \emph{$\varGamma_{\rm C}$-formula} $I$ is a conjunction of the form
		\begin{align}\label{eq:CSP-C-instance}
			I = \bigwedge_{i = 1}^{m} R_i(\xi^i_1,\dots,\xi^i_{r_i}),
		\end{align}
		where $m$ is a positive integer, each $R_i$ is an $r_i$-ary relation from $\varGamma$, 
		and the $\xi^i_{j}$ are (not necessarily distinct) variables or elements of $D$ (also called \emph{constants}).
		
		We denote the set of variables that occur in $I$ by $\var(I)$.
		An assignment $\assi: \var(I) \to D$ \emph{satisfies} $I$ or is a \emph{solution} of $I$ if for all $i \in [m]$ it holds that $(\assi(\xi^i_1),\dots, \assi(\xi^i_{r_i})) \in R_i$, where we define $\assi(d) = d$ for every $d \in D$.
	\end{definition}
	
	\ifconference
\else
	Note that each formula defines the relation of its satisfying assignments (or solutions).
	Indeed, we assume that $\var(I) = \{x_1,\dots,x_n\}$ for a $\varGamma$-formula (resp., $\varGamma_{\rm C}$-formula) $I$, 
	and identify an assignment $\assi$ with a tuple of $\bs{t}$ in $D^{n}$ such that $t_j = \assi(x_j)$ for all $j \in [n]$.
	We also denote by $s(I)$ the set of solutions of $I$ and regard it as an $n$-ary relation, i.e., $s(I) \subseteq D^n$.
	We say that $s(I)$ is \emph{expressed} by a $\varGamma$-formula (resp., $\varGamma_{\rm C}$-formula).
	\fi 
	
	\ifconference
	\else
	A \emph{conjunctive normal form (CNF)} formula is a propositional formula of the form $C_1 \wedge \dots \wedge C_m$ ($m \ge 1$), where each $C_i$ is a \emph{clause}, that is, a finite disjunction of \emph{literals} (variables or negated variables).
	A \emph{$k$-CNF formula} ($k \ge 1$) is a CNF formula where each $C_i$ has at most $k$ literals.
	The Boolean satisfiability problem (SAT) is a problem to determine if a given CNF formula is satisfiable, and 
	$k$-SAT is a problem to determine if a given $k$-CNF formula is satisfiable.
	\fi 
	
	\begin{example}\label{ex:2-CNF}
		Let $D = \{0,1\}$ and $\varGamma_{\rm 2SAT} = \{ R_{ij} \mid i,j \in \{0,1\} \}$, where $R_{ij} = \{0,1\}^2 \setminus \{(i,j)\}$.
		Consider an instance (2-CNF) $\varphi$ of 2-SAT $\varphi(x_1,x_2,x_3) = (x_1 \vee x_2) \wedge \overline{x_2} \wedge x_3$.
		Then $\varphi$ can be represented as a $\varGamma_{\rm C}$-formula
		$I_{\varphi} = R_{00}(x_1,x_2) \wedge R_{01}(0,x_2) \wedge R_{00}(x_3,x_3)$.
		The set $s(I_{\varphi})=\{(1,0,1)\} \subseteq \{0,1\}^3$.
	\end{example}
	
	\begin{definition}[CSP]
		\ifconference
		\else
		Let $\varGamma$ be a constraint language on $D$.
		
		The \emph{constraint satisfaction problem (CSP) on $\varGamma$}, denoted by $\csp(\varGamma)$, consists of the instances $I$ of the form \eqref{eq:CSP-instance}.
		
		The \emph{constraint satisfaction problem on $\varGamma$ with constants}, denoted by $\cspc(\varGamma)$, consists of the instances $I$ of the form \eqref{eq:CSP-C-instance}.
		
		An instance $I$ is a yes instance if it has a solution and a no instance otherwise.
		\fi 
	\end{definition}
	
	\begin{example}
		\ifconference
\else
		The $k$-SAT problem is a CSP over the Boolean domain $\{0,1\}$.
		In addition, the graph $k$-coloring problem is equivalent to $\csp(\{\neq_D \})$, where $D = \{0,1,\dots,k-1\}$.
		\fi 
	\end{example}
	
	
	Now, we define the reconfiguration CSP (RCSP).
	
	\begin{definition}[Solution graph]
		For a relation $R \subseteq D^r$, 
		the \emph{solution graph $G(R) = (V(R), E(R))$ associated with $R$} is an undirected graph defined as follows.
		Its vertex set is $R$, i.e., $V(R)= R$.
		Moreover, for $\bs{x},\bs{y} \in R$, we have $\{\bs{x}, \bs{y}\} \in E(R)$ if and only if $\dist(\bs{x},\bs{y}) = 1$, 
		where $\dist(\bs{x},\bs{y})= |\{ j \mid x_j \neq y_j \}|$ is the Hamming distance of $\bs{x}$ and $\bs{y}$.
		
		For a CSP instance $I$, the solution graph $G(s(I))$ of $s(I)$ is denoted by $G(I)$ for brevity.
	\end{definition}
	
	\begin{definition}[RCSP]
		\ifconference
\else
		Let $\varGamma$ be a constraint language on $D$.
		The \emph{reconfiguration constraint satisfaction problem (RCSP) on $\varGamma$}, denoted by $\rcsp(\varGamma)$, consists of the instances of the form $(I, \bs{s}, \bs{t})$, 
		where $I$ is an instance of $\csp(\varGamma)$ and $\bs{s}, \bs{t}$ are solutions of $I$.
		
		The \emph{RCSP with constants on $\varGamma$}, denoted by $\rcspc(\varGamma)$, is defined analogously.

		An instance $(I, \bs{s}, \bs{t})$ is a yes instance if $\bs{s}$ and $\bs{t}$ are in the same connected component in $G(I)$, and a no instance otherwise.
		\fi 
	\end{definition}
	
	For CSPs, the computational complexity of $\mathrm{CSP}(\varGamma)$ remains unchanged when $\varGamma$ is replaced by its core (i.e., the unique up to isomorphism induced substructure of $\varGamma$ with no non-surjective endomorphisms) and all constant relations are added. Therefore, considering CSPs with constants does not impose any restriction in the context of complexity classification. In contrast, for RCSPs, retracting to the core may alter the yes/no status of instances, so it is not clear whether considering RCSPs with constants is without loss of generality.
	
	
	\ifconference
	\else
	\subsection{Polymorphism and logical expression (reduction)}
	\label{subsec:pol}
	
	In the algebraic approach to the CSP, 
	the computational complexity of the CSP is characterized using polymorphisms, which capture \emph{symmetries} of the problems.
	
	\begin{definition}[Operation and polymorphism]
		A $k$-ary operation on $D$ is a mapping $D^k \to D$.
		A $k$-ary operation $f$ is a \emph{polymorphism} of an $r$-ary relation $R \subseteq D^r$ if $f$ applied componentwise to any $k$ elements of $R$ gives an element of $R$.
		In more detail, whenever $\bs{x}^{(1)},\dots,\bs{x}^{(k)}$ are in $R$, 
		$f(\bs{x}^{(1)},\dots,\bs{x}^{(k)})$ is also in $R$, 
		where $\bs{x}^{(i)} = (x_1^{(i)},\dots , x_r^{(i)})$ for each $i \in [k]$ and 
		\begin{align*}
		f(\bs{x}^{(1)},\dots , \bs{x}^{(k)}) = \left(f(x_1^{(1)},\dots, x_1^{(k)}),\dots , f(x_r^{(1)},\dots, x_r^{(k)}) \right)\in D^r.
		\end{align*}
		
		If $f$ is a polymorphism of $R$, $R$ is \emph{invariant} under $f$.
		
		For a set $\Gamma$ of relations, $f$ is called a polymorphism of $\Gamma$ if it is a polymorphism of every relation in $\Gamma$.
		In this case, $\Gamma$ is \emph{invariant} under $f$.
	\end{definition}
	
	\begin{example}
		A ternary (i.e., 3-ary) operation $M:D^3 \to D$ is called \emph{majority} 
		if for all $x,y \in D$, it holds that $M(x,x,y) = M(x,y,x) = M(y,x,x) = x$.
        The unique majority operation on $\{0,1\}$ is denoted by $\maj$.
		
		A ternary operation $M:D^3 \to D$ is called \emph{Maltsev} 
		if for all $x,y \in D$, it holds that $M(x,y,y) = M(y,y,x) = x$.
		A Maltsev operation is a generalization of the affine operation $f(x,y,z) = x - y + z$, which preserves the solution space of linear equations.
		
		
		A \emph{Taylor} operation is a $k$-ary ($k\ge 2$) operation $f$ such that, for each $1 \le i \le k$, 
		$f$ satisfies an identity of the form 
		\begin{align*}
			f(z_{i,1},\dots, z_{i,i-1},x,z_{i,i+1},\dots,z_{i,k}) = f(z'_{i,1},\dots, z'_{i,i-1},y,z'_{i,i+1},\dots,z'_{i,k})
		\end{align*}
		where all $z_{i,j},z'_{i,j}$ are in $\{x,y\}$.
	\end{example}
	
	\begin{definition}[Pol and Inv]
		For a set $\varGamma$ of relations on $D$, 
		the set $\pol(\varGamma)$ denotes the set of operations that are a polymorphism of $\varGamma$.
		For a set $F$ of operations on $D$, 
		the set $\inv(F)$ denotes the set of relations that are invariant under every operation in $F$.
	\end{definition}
	
	As stated below, 
	the relations that are invariant under polymorphisms of a set $\varGamma$ of relations on $D$ coincide with those relations expressible by a formula using relations in $\varGamma \cup \{\Delta_D\} \cup \{ \emptyset^{(1)} \}$ and existential quantifiers.
	
	\begin{definition}
		Let $\varGamma$ be a set of relations on $D$.
		We denote by $\langle \varGamma \rangle_{\exists,\wedge,=,\false}$ the set of relations expressible by $(\varGamma \cup \{\Delta_D\} \cup \{\emptyset^{(1)}\})$-formulas with additional existentially quantified variables.
		Explicitly, a relation $R \in \langle \varGamma \rangle_{\exists,\wedge,=,\false}$ can be represented in the form
		\begin{align}
			R = \exists x^{m+1}_{1} \cdots x^{m+1}_{r_{m+1}}\bigwedge_{i = 1}^{m} R_i(x^i_1,\dots,x^i_{r_i}),
		\end{align}
		where each $R_i$ is an $r_i$-ary relation from 
		$(\varGamma \cup \{\Delta_D\} \cup \{ \emptyset^{(1)}\})$, 
		and the variables $x^i_j$ are not necessarily distinct.
		In this case, we say that $R$ has a \emph{primitive positive definition (pp-definition) over $\varGamma$}, or equivalently, that $R$ is \emph{pp-definable from $\varGamma$}.
	\end{definition}

	The following is a well-known fact in the algebraic approach to the CSP.
	
	\begin{proposition}[{\citealp{Gei68}}]
		\label{prop:Pol-logical-characterization}
		Let $\Gamma$ be a set of relations on $D$. Then
		\[
		\langle \Gamma \rangle_{\exists,\wedge,=,\false}
		=
		\inv(\pol(\Gamma)).
		\]
	\end{proposition}
	

	
	Moreover, 
	the inclusion relationship between polymorphisms induces reductions between the corresponding CSPs.
	Consequently, the set of polymorphisms characterizes the computational complexity of the CSP on a given constraint language $\varGamma$.
	
	\begin{theorem}[{\citealp[Corollary 4.11]{Jea98}}]
		Let $\varGamma_1$ and $\varGamma_2$ be constraint languages on $D$.
		Assume that $\pol(\varGamma_1) \subseteq \pol(\varGamma_2)$.
		Then $\csp(\varGamma_2)$ is polynomial-time reducible to $\csp(\varGamma_1)$.
	\end{theorem}
	
	In fact, the following dichotomy theorem, which characterizes the computational complexity of $\csp(\varGamma)$ in terms of its set of polymorphisms, is well known.
	
	\begin{theorem}[\citealp{Bul17,Zhu17,Zhu20}]
		Let $\varGamma$ be a constraint language on $D$.
		If $\pol(\varGamma)$ contains a Taylor operation, then $\csp(\varGamma)$ is polynomial-time solvable; otherwise, $\csp(\varGamma)$ is NP-complete.
	\end{theorem}

	The following lemma states that the polymorphisms of a relation $R$ are inherited by the connected components of $G(R)$.
	An operation $f: D^k \rightarrow D$ is called {\it idempotent} if 
	$f(x, x, \dots, x) = x$ holds for any $x \in D$.
	
	\begin{lemma}[{\citealp[Lemma 2.1]{KiS21}}]\label{lem:connected_component_closedness}
		If a relation $R \subseteq D^n$ is invariant under an idempotent operation $f: D^k \rightarrow D$, 
		then every connected component of $G(R)$ is also invariant under $f$.
	\end{lemma}
	
	\fi 
	\subsection{Partial polymorphism and logical expression (reduction)}
	\label{subsec:ppol}
	\ifconference
\else
	In this subsection, parallel to \Cref{subsec:pol}, 
	we show that the computational complexity of the RCSP is characterized using partial operations.
	\fi 
	
	\begin{definition}[Partial operation]
		A $k$-ary \emph{partial operation} $f: D^k \pto D$ on $D$ is a mapping $f:D' \to D$ for some $D' \subseteq D^k$.
		Here, $D'$ is called the domain of $f$ and is denoted by $\dom(f)$.
		If $\bs{x} \in D'$, then we say that $f(\bs{x})$ is \emph{defined} and otherwise (that is, if $\bs{x} \notin D'$) \emph{undefined}.
		If $\dom(f) = D^k$, then $f$ is a \emph{total} operation.
	\end{definition}

    We present an example of a partial operation, namely the partial
Maltsev operation, as given in \citet[Example~3.7]{LW20} and
\citet[Definition~3.19]{LW21}.

\begin{example}
\label{Ex:partial-Maltsev}
    The partial Maltsev operation $\Mal_{p}:D^3 \pto D$ on $D$ is
    defined as follows.
    Its domain is
    \[
        \dom(\Mal_{p})
        =
        \{(x,y,y) \mid x,y \in D\}
        \cup
        \{(y,y,x) \mid x,y \in D\}
        \subseteq D^3.
    \]
    For all $x,y \in D$, it satisfies
    \[
        \Mal_{p}(x,y,y)=\Mal_{p}(y,y,x)=x.
    \]
\end{example}

	\begin{definition}\label{def:partial-order-on-partial-operations}
		Let $f$ and $g$ be two partial operations on $D$ with the same arity. 
		We say that $f$ is a \emph{subfunction} of $g$ if $\dom(f) \subseteq \dom(g)$ 
		and $f(\bs{x}) = g(\bs{x})$ for every $\bs{x} \in \operatorname{dom}(f)$. 
		We write $f \le g$ to denote this relation.
	\end{definition}

	The partial Maltsev operation is a partial version (more precisely, a subfunction) of a Maltsev operation, where Maltsev operations have been extensively studied in the algebraic approach to constraint satisfaction problems (CSPs).
	
	A partial operation $f$ on $D$ is called \emph{idempotent} if 
	$f(x,\dots,x)=x$ holds for every $x \in D$.
	Note that the partial Maltsev operation is idempotent.
	
	Now, we define the notion of partial polymorphism (e.g., \citealp{LW21}), a key concept in the algebraic approach to RCSPs. Intuitively, a partial operation is said to be a partial polymorphism of a constraint language $\varGamma$ if, whenever it is defined on several tuples from a relation in $\varGamma$, the result of applying the operation to those tuples also belongs to the same relation.
	
	For a $k$-ary partial operation $f$ on $D$ and tuples $\bs{x}^{(1)},\dots,\bs{x}^{(k)}$ in $D^r$, define $f(\bs{x}^{(1)},\dots , \bs{x}^{(k)})$ as $f(\bs{x}^{(1)},\dots , \bs{x}^{(k)}) = \left(f(x_1^{(1)},\dots, x_1^{(k)}),\dots , f(x_r^{(1)},\dots, x_r^{(k)}) \right)\in D^r$, 
	where $\bs{x}^{(i)} = (x_1^{(i)},\dots , x_r^{(i)})$ for each $i \in [k]$.
	If $f(x_j^{(1)},\dots, x_j^{(k)})$ is defined for all $j \in [r]$, then we say $f(\bs{x}^{(1)},\dots , \bs{x}^{(k)})$ is \emph{defined}; otherwise it is \emph{undefined}.
	
	\begin{definition}[Partial polymorphism]
		A $k$-ary partial operation $f$ on $D$ is a \emph{partial polymorphism} of an $r$-ary relation $R \subseteq D^r$ if for any $\bs{x}^{(1)},\dots,\bs{x}^{(k)} \in R$ such that $f(\bs{x}^{(1)},\dots , \bs{x}^{(k)})$ is defined, we have $f(\bs{x}^{(1)},\dots,\bs{x}^{(k)})\in R$.
		
		If $f$ is a partial polymorphism of $R$, $R$ is \emph{invariant} under $f$.
		
		For a set $\Gamma$ of relations, $f$ is called a partial polymorphism of $\Gamma$ if it is a polymorphism of every relation in $\Gamma$.
		In this case, $\Gamma$ is \emph{invariant} under $f$.
	\end{definition}
	
	\begin{definition}[pPol and Inv]
		Let $\varGamma$ be a set of relations on $D$.
		The set $\ppol(\varGamma)$ denotes the set of partial operations that are a partial polymorphism of $\varGamma$, i.e., 
		\begin{align*}
			\ppol(\varGamma) = \{ f \mid \text{for every $R \in \varGamma$, $f$ is a partial polymorphism of $R$} \}.
		\end{align*}
		
		Let $F$ be a set of partial operations on $D$.
		The set $\inv(F)$ denotes the set of relations that are invariant under every partial operation in $F$, i.e., 
		\begin{align*}
			\inv(F) = \{ R \mid \text{for every $f \in F$, $R$ is invariant under $f$} \}.
		\end{align*}
	\end{definition}
	
	
	The fact that the set of partial polymorphisms is closed under taking subfunctions is well established in the literature and follows directly from the definition.
	
	\begin{lemma}\label{obs:ppol-strong-clone}
		Let $\varGamma$ be a set of relations on $D$.
		\ifconference
\else
		If $g \in \ppol(\varGamma)$ and $f \le g$, then $f \in \ppol(\varGamma)$.
		\fi
	\end{lemma}
	
	\ifconference
	\else
	Since any operation is a partial operation, for any set $\varGamma$ of relations on $D$, 
	we have $\pol(\varGamma) \subseteq \ppol(\varGamma)$.
	\fi

	\ifconference
\else
	A result similar to \cref{prop:Pol-logical-characterization} is known.
	\fi 
	
	\begin{definition}
		Let $\varGamma$ be a set of relations on $D$. We denote by $\langle \varGamma \rangle_{\wedge,=,\false}$ the set of relations expressible by $(\varGamma \cup \{\Delta_D\} \cup \{\emptyset^{(1)}\})$-formulas. For any relation $R \in \langle \varGamma \rangle_{\wedge,=,\false}$, we say that $R$ has a \emph{quantifier-free primitive positive definition (qfpp-definition) over $\varGamma$}, or equivalently, that $R$ is \emph{qfpp-definable from $\varGamma$}.
	\end{definition}

	\begin{proposition}[{\citealp{Gei68,Rom81}}]
		\label{prop:pPol-logical-characterization}
For any set $\varGamma$ of relations on $D$,
\[
\inv(\ppol(\varGamma))
=
\langle \varGamma \rangle_{\wedge,=,\false}.
\]
	\end{proposition}
	
	\ifconference
\else
	Consequently, $\varGamma$ can be characterized by partial operations if and only if $\varGamma = \langle \varGamma \rangle_{\wedge,=,\false}$.
	Moreover, the partial polymorphisms of a constraint language $\varGamma$ are inherited by the set of solutions of an instance of $\csp(\varGamma)$ as shown below.
	
	\begin{corollary}\label{cor:pPol-inherited-by-CSP-instance}
		Let $\varGamma$ be a constraint language on $D$.
		Let $I$ be an instance of $\csp(\varGamma)$ and $f$ be a partial polymorphism of $\varGamma$.
		Then $s(I)$ is invariant under $f$.
	\end{corollary}
	\begin{proof}
		This follows from~\cref{prop:pPol-logical-characterization}, 
		since $s(I)$ can be represented as a $\varGamma$-formula.
	\end{proof}
	\fi 

	The following theorem shows that the inclusion relation of partial polymorphisms leads to a reduction between RCSPs provided that they have the equality relation.
	This allows us to characterize the computational complexity of RCSPs having the equality relation by the set of partial polymorphisms.
	The following proof is similar to that of Theorem 10 in~\citep{JLNZ17}. 
	
	\ifconference
\else
	\begin{theorem}
		\label{thm:pPol-capture-reduction}
		Let $\varGamma_1$ and $\varGamma_2$ be constraint languages on $D$.
		Assume that the equality relation $\Delta_D$ is contained in $\varGamma_1$
		and that $\ppol(\varGamma_1) \subseteq \ppol(\varGamma_2)$.
		Then $\rcsp(\varGamma_2)$ is polynomial-time reducible to $\rcsp(\varGamma_1)$.
	\end{theorem}
	\begin{proof}
		Given an instance $(I,\bs{s},\bs{t})$ of $\rcsp(\varGamma_2)$ with $n$ variables, we transform it into an equivalent instance $(I',\bs{s}',\bs{t}')$ of $\rcsp(\varGamma_1)$.
		From \cref{prop:pPol-logical-characterization}, 
		every constraint $R(x_1,...,x_r)$ in $I$ can be replaced with constraints
		\begin{align}
			R_1(x_{11},\dots,x_{1r_1})\wedge \dots \wedge R_\ell(x_{\ell 1},\dots,x_{\ell r_\ell}),
		\end{align}
		where $R_1,\dots,R_\ell \in \varGamma_1$ and $x_{11},\dots,x_{\ell r_\ell} \in \{ x_1,\dots, x_r\}$.
		The resulting instance $I'$ has the same set of solutions as that of $I$.
		By setting $\bs{s}'=\bs{s}$ and $\bs{t}'=\bs{t}$ , 
		we have that $(I,\bs{s},\bs{t})$ is a yes instance if and only if $(I',\bs{s}',\bs{t}')$ is a yes instance.
		Since each constraint in $I$ is replaced in constant time, the above reduction can be done in polynomial time.
	\end{proof}
	\fi 
	
	\begin{remark}
		As one reviewer pointed out, our framework can alternatively be formulated in terms of \emph{multifunctions}, i.e., mappings
		$f \colon D^k \rightarrow 2^D$
		that assign a (possibly empty) set of outputs to each input tuple. Partial functions can then be viewed as a special case of multifunctions in which the output set has size at most one for every input tuple. A \emph{multi-polymorphism} of a relation is a multifunction that preserves the relation in the natural coordinatewise sense: whenever tuples belong to the relation, every choice of outputs produced by the multifunction also yields a tuple in the relation. Under this viewpoint, the classical Geiger correspondence~\citep{Gei68} implies that one can characterize the closure under formulas built using conjunction ($\wedge$), the empty relation ($\emptyset$), and the full unary relation $D$, without introducing the equality relation. Consequently, inclusion of multi-polymorphism sets yields reductions between the corresponding RCSPs. Nevertheless, for our analysis of existing complexity results for Boolean RCSPs---which constitutes one of the main contributions of this paper---the framework of partial polymorphisms is more convenient and better suited to the arguments. Therefore, we focus primarily on partial polymorphisms throughout the paper.
	\end{remark}
		
	\ifconference
	\else
	It is easily seen from definition that for every constraint language $\varGamma$ on $D$, 
	$\cspc(\varGamma)$ and $\csp(\varGamma \cup \{ C_d \mid d \in D \})$ (resp., $\rcspc(\varGamma)$ and $\rcsp(\varGamma \cup \{ C_d \mid d \in D \})$) are polynomial-time reducible to each other.

	\begin{lemma}
		\label{lem:idempotent-constants}
		Let $f$ be an idempotent partial operation.
		Suppose that we prove the following proposition:
		for any constraint language $\varGamma$ that is invariant under $f$, $\csp(\varGamma)$ (resp., $\rcsp(\varGamma)$) is polynomial-time solvable.
		Then it also holds that 
		for any constraint language $\varGamma$ that is invariant under $f$, $\cspc(\varGamma)$ (resp., $\rcspc(\varGamma)$) is polynomial-time solvable.
	\end{lemma}
	
	\begin{proof}
		Suppose that for any constraint language $\varGamma$ that is invariant under $f$, $\csp(\varGamma)$ (resp., $\rcsp(\varGamma)$) is polynomial-time solvable.
		
		Now, assume that a constraint language $\varGamma'$ is invariant under $f$.
		It follows that $\varGamma' \cup \{ C_d \mid d \in D \}$ is also invariant under $f$.
		Hence, $\csp(\varGamma' \cup \{ C_d \mid d \in D \})$ (resp., $\rcspc(\varGamma'\cup \{ C_d \mid d \in D \})$ is polynomial-time solvable.
		
		Since $\cspc(\varGamma')$ and $\csp(\varGamma' \cup \{ C_d \mid d \in D \})$ (resp., $\rcspc(\varGamma')$ and $\rcsp(\varGamma' \cup \{ C_d \mid d \in D \})$) are polynomial-time reducible to each other as observed above, 
		$\cspc(\varGamma')$ (resp., $\rcspc(\varGamma')$) is also polynomial-time solvable.
	\end{proof}
	
	From the above lemma, 
	from now on, when dealing with constraint languages that are invariant under an idempotent partial operation, 
	we will not explicitly write the difference between $\csp$ and $\cspc$ (resp., $\rcsp$ and $\rcspc$).
	\fi 
	
	\subsection{Known results for the Boolean RCSP}
	\label{subsec:results-for-Boolean-RCSP}
	
	We summarize the results by \citet{GKMP09} and \citet{Sch14} for the RCSP on the Boolean domain, i.e., the case of $D = \{0,1\}$.
	
	For an $r$-ary relation $R$ on $D$ and $0 < k < r$, we can define an $(r-k)$-ary relation $R'(x_1,\dots, x_{r-k})=R(\xi_1,\dots,\xi_n)$, 
	where each $\xi_j \in \{ x_1,\dots,x_{r-k} \} \cup D$.
	If each variable $x_j$ occurs at most once in $(\xi_1,\dots,\xi_n)$, 
	we say that $R'$ is obtained from $R$ by \emph{substitution of constants}.
	If each $\xi_j$ is a variable, we say that $R'$ is obtained from $R$ by \emph{identification of variables}.
	
	\begin{example}
		Recall the example in \Cref{ex:2-CNF}.
		There, $R'_{01}(x_2):=R_{01}(0,x_2)$ is obtained from $R_{01}$ by substitution of constants.
		Moreover, $R'_{00}(x_3):=R_{00}(x_3,x_3)$ is obtained from $R_{00}$ by identification of variables.
	\end{example}
	
	\begin{definition}
		Let $R$ be a relation on $\{0,1\}$.
		\begin{itemize}
			\item $R$ is \emph{OR-free} (resp., \emph{NAND-free}) if the binary relation ${\rm OR} = \{ (0,1), (1,0), (1,1) \}$ (resp., ${\rm NAND} = \{ (0,0), (0,1), (1,0) \}$) cannot be obtained from $R$ by substitution of constants.
			\item $R$ is \emph{safely OR-free} (resp., \emph{safely NAND-free}) if $R$ and every relation $R'$ obtained from $R$ by identification of variables is OR-free (resp., NAND-free).
		\end{itemize}
		We denote the set of safely OR-free (resp., safely NAND-free) relations by $\varGamma_{\sor}$ (resp., $\varGamma_{\snand}$).
	\end{definition}
	
	\begin{definition}
		Let $R$ be a relation on $\{0,1\}$.
		\begin{itemize}
			\item $R$ is \emph{bijunctive} if it is the set of solutions of a 2-CNF-formula.
			\item $R$ is \emph{componentwise bijunctive} if every connected component $G(R)$ is a bijunctive relation.
			\item $R$ is \emph{safely componentwise bijunctive} if $R$ and every relation $R'$ obtained from $R$ by identification of variables is componentwise bijunctive.
		\end{itemize}
		We denote the set of safely componentwise bijunctive relations by $\varGamma_{\scb}$.
	\end{definition}
	
	\ifconference
	\else
	It is known that bijunctive relations can be characterized by invariance under an operation.
	
	\begin{lemma}[{\citealp{Sch78}}]
		\label{lem:bijunctive<=>majority-invariant}
		Let $R$ be a relation on $\{0,1\}$.
		Then $R$ is bijunctive if and only if it is invariant under 
		the majority operation on $\{0,1\}$.
	\end{lemma}
	\fi 
	
	Now, we introduce the relations investigated in \citep{GKMP09,Sch14}.
	
	\begin{definition}
		A set $\varGamma$ of relations on $\{0,1\}$ is \emph{tight} (resp., \emph{safely tight}) if at least one of the following conditions holds:
		\begin{itemize}
			\item every relation in $\varGamma$ is componentwise bijunctive (resp., safely componentwise bijunctive).
			\item every relation in $\varGamma$ is OR-free (resp., safely OR-free).
			\item every relation in $\varGamma$ is NAND-free (resp., safely NAND-free).
		\end{itemize}
	\end{definition}
	
	The following dichotomy for the computational complexity of $\rcspc$ on the Boolean domain is essentially shown by \citet{GKMP09}.
	However, the result is later corrected by \citet{Sch14} by adding ``safely'' to the condition for polynomial-time solvability.
	
	\begin{theorem}[{\citealp[Theorem 2.9]{GKMP09}, and \citealp{Sch14}}]
		Let $\varGamma$ be a constraint language on $\{0,1\}$.
		If $\varGamma$ is safely tight, then $\rcspc(\varGamma)$ is in P; 
		otherwise, $\rcspc(\varGamma)$ is PSPACE-complete.
	\end{theorem}

	\section{Safely OR-free (NAND-free) relations and their extension}
	\label{sec:SOR}
	
	
	In this section, we characterize the sets of safely OR-free and safely NAND-free relations, denoted by $\varGamma_{\sor}$ and $\varGamma_{\snand}$ respectively, using a single partial operation on $\{0,1\}$. 
	We then extend this operation to larger domains, yielding a class of RCSPs solvable in polynomial time. 
	Finally, in \cref{subsec:Maltsev-relations}, we show that this class encompasses several known CSP and graph homomorphism cases, including instances where the tractability of RCSPs was previously unknown.
	
	\subsection{Ordered partial Maltsev operation and characterization of \texorpdfstring{$\varGamma_{\sor}$}{Γₛₒᵣ}}
	
	First, we introduce a new partial operation based on the Maltsev operation.
	
	\begin{definition}
		Let $D$ be a finite set equipped with a total order $\le$.
		The \emph{$(D,\le)$-Maltsev} operation $M_{D,\le}$ is the ternary partial operation $M_{D,\le}:D^3 \pto D$ that satisfies 
		$M_{D,\le}(x,y,y) = M_{D,\le}(y,y,x) = x$ for any $x,y \in D$ with $x \le y$ and $M_{D,\le}(x,y,z)$ being undefined for the other inputs.
		
		We call a partial operation \emph{ordered partial Maltsev} if it coincides with $M_{D,\le}$ for some totally ordered finite set $(D,\le)$.
	\end{definition}
	
	By definition, an ordered partial Maltsev operation is a subfunction of the partial Maltsev operation, defined in \Cref{Ex:partial-Maltsev}.
	Note that by definition ordered partial Maltsev operations are idempotent.
	\begin{example}\label{ex:0<1-partial-Maltsev-op}
		When $D = \{0,1\}$ and $0 < 1$, 
		the \emph{$(D,\le)$-Maltsev} operation $M_{D,\le}$ is a ternary partial operation $M_{D,\le}:\{0,1\}^3 \pto \{0,1\}$ such that 
		$M_{D,\le}(0,0,0) = M_{D,\le}(0,1,1) = M_{D,\le}(1,1,0) = 0$, $M_{D,\le}(1,1,1)=1$, and undefined for the other inputs.
	\end{example}
	
	\begin{example}\label{ex:1<0-partial-Maltsev-op}
		When $D = \{0,1\}$ and $1 < 0$, 
		the \emph{$(D,\le)$-Maltsev} operation $M_{D,\le}$ is a ternary partial operation $M_{D,\le}:\{0,1\}^3 \pto \{0,1\}$ such that 
		$M_{D,\le}(0,0,0) = 0$, $M_{D,\le}(1,0,0) = M_{D,\le}(0,0,1) = M_{D,\le}(1,1,1)=1$, and undefined for the other inputs.
	\end{example}
	
	We will show that the partial operations in~\cref{ex:0<1-partial-Maltsev-op,ex:1<0-partial-Maltsev-op} characterize the sets of safely OR-free and safely NAND-free relations, respectively.
	
	\begin{lemma}\label{thm:OR-free=partial-ordered-Maltsev}
		Let $D = \{0,1\}$ and $0 < 1$. 
		Let $r$ be a positive integer and $R$ be an $r$-ary relation $R \subseteq \{0,1\}^r$.
		Then, $R$ is safely OR-free if and only if 
		it is invariant under the $(D,\le)$-Maltsev operation.
	\end{lemma}
	
	\ifconference
\else
	\begin{proof}
		The proof, especially the construction of the OR relation in the ``if'' part, is based on a variable identification argument analogous to that found in \citep[Lemma 8.1]{Marx2005}.
		
		We first show the only-if part by contradiction.
		Assume that $R$ is invariant under $M_{D,\le}$ but not safely OR-free.
		Then there exists $\xi \in \{ x_1, x_2, 0, 1 \}^r$ such that $R'(x_1,x_2):=R(\xi_1, \dots, \xi_r)$ is OR($=\{(0,1),(1,0),(1,1)\}$).
		It follows that, up to permutation, $R$ contains tuples 
		\begin{align}
			&\bs{t}^1=(\overbrace{0,\dots,0}^{\xi_i=x_1},\overbrace{1,\dots,1}^{\xi_i=x_2},\overbrace{0,\dots,0}^{\xi_i=0},\overbrace{1,\dots,1}^{\xi_i=1})\ (\text{which corresponds to $(0,1)$ in $R'$})\\
			&\bs{t}^2=(\overbrace{1,\dots,1}^{\xi_i=x_1},\overbrace{1,\dots,1}^{\xi_i=x_2},\overbrace{0,\dots,0}^{\xi_i=0},\overbrace{1,\dots,1}^{\xi_i=1})\ (\text{which corresponds to $(1,1)$ in $R'$})\\
			&\bs{t}^3=(\overbrace{1,\dots,1}^{\xi_i=x_1},\overbrace{0,\dots,0}^{\xi_i=x_2},\overbrace{0,\dots,0}^{\xi_i=0},\overbrace{1,\dots,1}^{\xi_i=1})\ (\text{which corresponds to $(1,0)$ in $R'$}).
		\end{align}
		Applying $M_{D,\le}$ to $\bs{t}^1,\bs{t}^2,\bs{t}^3$ yields 
		\begin{align}
			(\overbrace{0,\dots,0}^{\xi_i=x_1},\overbrace{0,\dots,0}^{\xi_i=x_2},\overbrace{0,\dots,0}^{\xi_i=0},\overbrace{1,\dots,1}^{\xi_i=1})\ (\text{which corresponds to $(0,0)$}).
		\end{align}
		Hence, $R'$ must also contain $(0,0)$, a contradiction.
		Therefore, $R$ is safely OR-free.
		
		We then show the if part.
		We show the contrapositive, i.e., assuming that $R$ is not invariant under $M_{D,\le}$, we show that $R$ is not safely OR-free.
		Assume that $R$ is not invariant under $M_{D,\le}$. 
		Then there exists $\bs{t}^1,\bs{t}^2,\bs{t}^3 \in R$ such that 
		$M_{D,\le}(\bs{t}^1,\bs{t}^2,\bs{t}^3)$ is defined and $M_{D,\le}(\bs{t}^1,\bs{t}^2,\bs{t}^3) \notin R$.
		Define sets of indices $I_{x_1}, I_{x_2}, I_{0}, I_{1} \subseteq [r]$ as 
		\begin{align}
			&I_{x_1} = \{ i \in [r] \mid t^1_{i} = 0 \wedge t^2_{i} = t^3_{i} = 1\} \\
			&I_{x_2} = \{ i \in [r] \mid t^1_{i} = t^2_{i} = 1 \wedge t^3_{i} = 0\} \\
			&I_{0} = \{ i \in [r] \mid t^1_{i} = t^2_{i} = t^3_{i} = 0\} \\
			&I_{1} = \{ i \in [r] \mid t^1_{i} = t^2_{i} = t^3_{i} = 1\}.
		\end{align}
		Then define $\xi \in \{ x_1, x_2, 0, 1 \}^r$ according to $I_{x_1}, I_{x_2}, I_{0}, I_{1}$, i.e., 
		\begin{align}
			\xi_i = 
			\begin{cases}
				x_1 & \text{if $i \in I_{x_1}$,} \\
				x_2 & \text{if $i \in I_{x_2}$,} \\
				0 & \text{if $i \in I_{0}$,} \\
				1 & \text{if $i \in I_{1}$.}
			\end{cases}
		\end{align}
		Now, define $R'(x_1,x_2)=R(\xi_1, \dots, \xi_r)$.
		Since $\bs{t}^1,\bs{t}^2,\bs{t}^3 \in R$, we have $(0,1),(1,0),(1,1) \in R'$.
		Moreover, since $M_{D,\le}(\bs{t}^1,\bs{t}^2,\bs{t}^3) \notin R$, we have $(0,0) \notin R'$.
		Namely, $R' = \{(0,1),(1,0),(1,1)\}$, which is the relation OR.
		Therefore, $R'$ is not safely OR-free.
		This completes the proof.
	\end{proof}
	\fi
	
	The following theorem is immediate from the above lemma.
	
	\begin{theorem}\label{thm:SOF<=>ordered-Maltsev}
		Let $D = \{0,1\}$ and $0 < 1$.
		Then $\varGamma_{\sor} = \inv(M_{D,\le})$.
	\end{theorem}
	
	
	Dually, we can show the following.

	\begin{theorem}\label{thm:SNF<=>ordered-Maltsev}
		Let $D = \{0,1\}$ and $1 < 0$.
		Then $\varGamma_{\snand} = \inv(M_{D,\le})$.
	\end{theorem}

	\subsection{Algorithm for RCSP invariant under an ordered partial Maltsev operation}\label{subsec:alg-for-partial-ordered-Maltsev}
	
	Now we extend the algorithmic result for $\rcspc(\varGamma_{\sor})$ by \citet{GKMP09} to the case where $\varGamma$ is invariant under an ordered partial Maltsev operation on an arbitrary finite domain.
	The algorithm is similar to the one for $\rcspc(\varGamma_{\sor})$ in~\citep{GKMP09}.
	First, we show that there exists a unique locally minimal solution for each connected component of the solution graph of any instance of 
	\ifconference
\else
	$\rcsp(\varGamma)$.
	\fi
	Here, a solution is \emph{locally minimal} in the solution graph if it has no smaller neighboring element in the solution graph.
	This shows that by greedily descending from any solution to its neighbors, 
	we can efficiently find the unique locally minimal solution of the connected component to which the solution belongs. 
	Thus, the reconfiguration problem can be solved by finding the unique locally minimal solution of the connected component to which each solution belongs for two given solutions and checking whether they coincide.

	Formally, for a totally ordered finite set $(D,\le)$ and a relation $R \subseteq D^r$, 
	$\bs{t} \in R$ is \emph{locally minimal} in $R$ 
	if for all $\bs{t}' \in D^r$ with $\dist(\bs{t}',\bs{t}) = 1$ and $\bs{t}' \le \bs{t}$ (i.e., $t'_i \le t_i$ for all $i \in [r]$) we have $\bs{t}' \notin R$.
	
	We first prove that the uniqueness of locally minimal solutions holds for each connected component of the solution graph in instances of $\rcsp$ with constraint languages invariant under ordered partial Maltsev operations, just as in the case of safely OR-free relations.
	
	
	\begin{lemma}\label{lem:partial-ordered-Maltsev=>unique-local-minima}
		Let $D$ be a finite set equipped with a total order $\le$.
		Let $\varGamma$ be a constraint language on $D$.
		Assume that $M_{D,\le} \in \ppol(\varGamma)$.
		Let $I$ be an instance of $\csp(\varGamma)$.
		Then each connected component of the solution graph $G(I)$ has a unique locally minimal element.
		It follows that every solution is connected to the unique locally minimal solution in the same component via a monotonically decreasing path.
	\end{lemma}
	\ifconference
\else
	\begin{proof}
		Let $C$ be a connected component of $G(I)$.
		Assume to the contrary that $C$ contains two different unique locally minimal elements $\bs{a}$ and $\bs{b}$.
		Choose any $(\bs{a},\bs{b})$-path $P$ in $G(I)$ such that the first place where some value decreases is closest to $\bs{a}$.
		Denote this path by $\bs{a} = \bs{u}^0 \rightarrow \bs{u}^1 \rightarrow \dots \rightarrow \bs{u}^\ell =\bs{b}$.
		Note that, for each $j \in [\ell]$, the solutions $\bs{u}^{j-1}$ and $\bs{u}^{j}$ differ in exactly one coordinate. Since $D$ is totally ordered, the two vectors are comparable; that is, either $\bs{u}^{j-1} \le \bs{u}^{j}$ or $\bs{u}^{j-1} \ge \bs{u}^{j}$.
		Let $\bs{u}^i$ be the solution just before the first place where some value decreases in the path.
		Namely, we have $\bs{u}^0 \le  \bs{u}^1 \le \dots \le \bs{u}^{i-1} \le \bs{u}^i \ge \bs{u}^{i+1}$.
		Then $\bs{u}^i \neq \bs{a}$ and $\bs{u}^i \neq \bs{b}$, since $\bs{a}$ and $\bs{b}$ are locally minimal.
		Now, the solutions $\bs{u}^{i-1}$ and $\bs{u}^{i+1}$ differ in exactly two variables, say, in $x_1$ and $x_2$. 
		So $(u^{i-1}_1 u^{i-1}_2, u^i_1 u^i_2, u^{i+1}_1 u^{i+1}_2) = (pq, rq, rs)$ for some $p,q,r,s \in D$ with $p < r$ and $q > s$.
		
		Let $\bs{u} = M_{D,\le}(\bs{u}^{i-1},\bs{u}^{i},\bs{u}^{i+1})$.
		Then it holds that $u_1 =p, u_2 = s$, and $u_j = u^{i}_j$ for $j > 2$. 
		Since $s(I)$ is invariant under $M_{D,\le}$ by~\cref{cor:pPol-inherited-by-CSP-instance}, 
		$\bs{u}$ is also a solution of $I$ and 
		the path $\bs{u}^0 \rightarrow \bs{u}^1 \rightarrow \dots \rightarrow \bs{u}^{i-1} \rightarrow \bs{u} \rightarrow \bs{u}^{i+1} \rightarrow \dots \rightarrow \bs{u}^\ell$ is also an $(\bs{a},\bs{b})$-path in $C$.
		However, this contradicts the way we chose the original path $P$, 
		since $\bs{u}^{i-1} \ge \bs{u}$.
		It follows that $C$ contains only one locally minimal element.
		
		The unique locally minimal solution in a component is its minimum solution,
		because starting from any other assignment, which is not locally minimal by the above argument, in the component, 
		it is possible to keep moving to neighbors that are smaller until it reaches to the locally minimal solution. 
		Therefore, there is a monotonically decreasing path from any solution to the minimum element in the connected component.
	\end{proof}
	\fi
	
	\begin{theorem}\label{thm:ordered-Maltsev=>RCSP-is-P}
		Let $\varGamma$ be a constraint language on $D$.
		Assume that for some total order $\le$ on $D$ we have $M_{D,\le} \in \ppol(\varGamma)$.
		Then $\rcspc(\varGamma)$ is solvable in ${\rm O}(n^2m|D|^2)$ time.
	\end{theorem}
	\begin{proof}
		Let $(I,\bs{s},\bs{t})$ be an instance of $\rcspc(\varGamma)$.
		From~\cref{lem:partial-ordered-Maltsev=>unique-local-minima} we can compute a unique minimal element $\bs{s}_{\min}$ (resp., $\bs{t}_{\min}$) of the connected component that contains $\bs{s}$ (resp, $\bs{t}$) by greedily decreasing the solution as far as possible.
		Determining if some value of a variable can be decreased is done in ${\rm O}(nm|D|)$ time.
		Since the value of each variable can be decreased at most $|D|-1$ times, 
		we can reach $\bs{s}_{\min}$ (resp., $\bs{t}_{\min}$) by decreasing a value of some variable at most $n(|D|-1)$ times.
		Therefore, $\bs{s}_{\min}$ (resp., $\bs{t}_{\min}$) can be obtained in ${\rm O}(n^2m|D|^2)$ time.
		We output yes if $\bs{s}_{\min} = \bs{t}_{\min}$ and no otherwise.
	\end{proof}
	
	The following is immediate from the proof of \cref{thm:ordered-Maltsev=>RCSP-is-P}.
	
	\begin{corollary}
		Let $\varGamma$ be a constraint language on $D$.
		Assume that for some total order $\le$ on $D$ we have $M_{D,\le} \in \ppol(\varGamma)$.
		Let $I$ be an instance of $\cspc(\varGamma)$.
		Then the diameter of each connected component of $G(I)$ is ${\rm O}(n|D|)$.
	\end{corollary}
	
	\ifconference
	\else
	In the following subsection, we will present some of the consequences of this result.
	\fi
	
	\subsection{Relations invariant under some ordered partial Maltsev operation}
	\label{subsec:Maltsev-relations}

	
	In this subsection, we enumerate relations that have been studied in the context of CSPs and are invariant under certain ordered partial Maltsev operations, which are summarized in \Cref{tab:Maltsev-invariant-relations}.
	
\begin{table}[t!]
    \centering
    \small
    \setlength{\tabcolsep}{3pt}
    \renewcommand{\arraystretch}{1.2}

    \begin{threeparttable}
        \caption{
            Relations or problems invariant under a Maltsev operation
            or one of its extensions.
            Results established in this paper are accompanied by
            theorem or proposition references.
        }
        \label{tab:Maltsev-invariant-relations}

        \begin{tabularx}{\linewidth}{
            |>{\centering\arraybackslash}m{1.25cm}
            |Y
            |Y
            |Y|
        }
            \hline

            &
            \makecell[c]{Ordered partial\\Maltsev}
            &
            \makecell[c]{Partial\\Maltsev}
            &
            Maltsev
            \\

            \hline

            \makecell[c]{General\\CSPs}
            &
            \makecell[c]{
                Min-closed\\
                {[Thm.~\ref{lem:min-closed=>ordered-Maltsev-closed}]}
            }
            &
            &
            \makecell[c]{
                Strongly rectangular\tnote{a}\\
                Lin.\ eqs.\ over $\mathrm{GF}(q)$\tnote{b}
            }
            \\

            \hline

            \makecell[c]{Boolean\\CSPs}
            &
            \makecell[c]{
                Safely OR-free\\                 {[Thm.~\ref{thm:SOF<=>ordered-Maltsev}]}\\
                Safely NAND-free\\
                {[Thm.~\ref{thm:SNF<=>ordered-Maltsev}]}
            }
            &
            \makecell[c]{
                Exact SAT\tnote{c}\\
                Subset Sum\tnote{c}
            }
            &
            \makecell[c]{
                Lin.\ eqs.\ over
                $\mathrm{GF}(2)$\tnote{d}
            }
            \\

            \hline

            Graphs
            &
            &
            \makecell[c]{
                Rectangular\\
                {[Thm.~\ref{thm:rectangular<=>partial-Maltsev}]}
            }
            &
            \makecell[c]{
                Circular clique $C_{6,3}$\\
                {[Prop.~\ref{prop:C_63-Maltsev}]}
            }
            \\

            \hline

            Digraphs
            &
            \makecell[c]{
                Trans.\ tourn.\\
                $r(\overrightarrow{K_n})$\\
                {[Thm.~\ref{thm:transitive-tournament-poMaltsev-invariant}]}
            }
            &
            \makecell[c]{
                Rectangular\\
                {[Thm.~\ref{thm:rectangular<=>partial-Maltsev}]}
            }
            &
            \makecell[c]{
                Totally rectangular\tnote{e}\\
                Graphs in Fig.~\ref{fig:cycle4}\\
                {[Prop.~\ref{prop:C_4^1C_4^2-Maltsev-invariant}]}
            }
            \\

            \hline
        \end{tabularx}

        \begin{tablenotes}[flushleft]
            \footnotesize

            \item[a]
            For strong rectangularity, see \citet{DR10}.

            \item[b]
            For systems of linear equations over finite fields,
            see \citet{JCG95}.

            \item[c]
            For Exact SAT and Subset Sum, see \citet{LW21}.

            \item[d]
            For systems of linear equations over $\mathrm{GF}(2)$,
            see \citet{Sch78}.

            \item[e]
            For total rectangularity, see \citet{CEJN15}.

            \item[]
            Here, ``trans.\ tourn.'' and ``lin.\ eqs.'' stand for
            ``transitive tournament'' and ``linear equations,''
            respectively.
            Moreover, $\mathrm{GF}(q)$ denotes the finite field
            with $q$ elements.
        \end{tablenotes}
    \end{threeparttable}
\end{table}

	\Cref{subsec:relation-Maltsev-invariant} discusses general relations, while \Cref{subsec:digraph-Maltsev-invariant} focuses on binary relations, namely digraphs.

	\subsubsection{Relations invariant under some ordered partial Maltsev operation}
	\label{subsec:relation-Maltsev-invariant}

	Note that for any Maltsev operation $M$ and the partial Maltsev operation $M_p$ on $D$, and any total order $\le$ on $D$, we have $M_{D,\le} \le M_p \le M$ (for the order relations between partial operations, please refer to \Cref{def:partial-order-on-partial-operations}).
	Thus, if a constraint language \( \varGamma \) is invariant under a Maltsev operation, it is also invariant under a partial Maltsev operation, and further under some ordered partial Maltsev operation by \Cref{obs:ppol-strong-clone}.
	
	First, we introduce relations invariant under a Maltsev operation.
	These relations appear in the study of counting CSP~\citep{DR10}.
	
	\begin{definition}[{\citealp{DR10}}]
		A binary relation $R \subseteq D^2$ is called \emph{rectangular} if 
		whenever $(x,y),(x',y),(x',y') \in R$ then also $(x,y') \in R$.\footnote{In \citep{DR10}, rectangularity is defined as ``whenever $(x,y),(x,y'),(x',y') \in R$ then also $(x',y) \in R$''; however, we adopt this particular definition here because it will be used later when presenting results for directed graphs. It should be noted that the invariance under (partial) operations remains unaffected by the choice of definition.}
		We call a set $\varGamma$ of relations \emph{strongly rectangular} if every binary relation $B \in \langle \varGamma \rangle_{\exists,\wedge,=}$ is rectangular.
	\end{definition}
	
	\begin{lemma}[{\citealp[Lemma 4]{DR10}}]
		Let $\varGamma$ be a set of relations.
		Then $\varGamma$ is strongly rectangular if and only if it is invariant under a Maltsev operation.
	\end{lemma}
	
	\begin{corollary}
		A strongly rectangular set of relations is invariant under some ordered partial Maltsev operation.
	\end{corollary}
	
	A well-known example of a relation that is invariant under a Maltsev operation is the solution set of a system of linear equations~\citep{JCG95}.
	
	Next, we introduce relations invariant under a partial Maltsev operation, especially on the Boolean domain $\{0,1\}$.
	These relations appear in the study of the fine-grained complexity of SAT, i.e., a study of fast exponential time algorithms for SAT.
	\citet{LW21} showed the solution sets of both the Exact SAT problem and the subset sum problem are invariant under the partial Maltsev operation on $\{0,1\}$.
	Here, Exact SAT is the SAT problem where the question is whether an input CNF has a satisfying assignment where every clause contains exactly one satisfied literal.
	Furthermore, the subset sum problem is defined as follows: given a multiset of integers and a target sum, the task is to determine whether there exists a subset of the integers whose sum is exactly equal to the target value.
	
	Finally, we show that any relation invariant under the minimum operation on a totally ordered set is also invariant under some ordered partial Maltsev operation.
	These relations are generalizations of Horn CNFs, which are intensively studied in the study of SAT~\citep{CH99}.
	
	\begin{lemma}
		\label{lem:min-closed=>ordered-Maltsev-closed}
		Let $\varGamma$ be a set of relations on $D$.
		Assume that for some total order $\le$ on $D$ the minimum operation $\min_{D,\le}$ is a polymorphism of $\varGamma$.
		Then we have $M_{D,\le} \in \ppol(\varGamma)$.
	\end{lemma}
	\begin{proof}
		Take any $R \in \varGamma$ and $\bs{t}_1, \bs{t}_2, \bs{t}_3 \in R$ such that $M_{D,\le}(\bs{t}_1, \bs{t}_2, \bs{t}_3)$ is defined.
		Define sets of indices $I_{=}, I_{<}, I_{>}$ as 
		\begin{align*}
			&I_= = \{i \mid t_{1i} = t_{2i} = t_{3i}\}\\
			&I_< = \{i \mid t_{1i} < t_{2i} = t_{3i}\}\\
			&I_> = \{i \mid t_{1i} = t_{2i} > t_{3i}\}.
		\end{align*}
		Then the $i$th component of $M_{D,\le}(\bs{t}_1, \bs{t}_2, \bs{t}_3)$ is 
		$t_{1i}$ if $i \in I_= \cup I_<$, and $t_{3i}$ if $i \in I_>$.
		It follows that $M_{D,\le}(\bs{t}_1, \bs{t}_2, \bs{t}_3) = \min_{D,\le}(\bs{t}_1, \bs{t}_3)$.
		Since $R$ is invariant under $\min_{D,\le}$, $M_{D,\le}(\bs{t}_1, \bs{t}_2, \bs{t}_3)$ is contained in $R$.
		Therefore, $R$ is also invariant under $M_{D,\le}$.
	\end{proof}

	\subsubsection{Digraphs invariant under some ordered partial Maltsev operation}
	\label{subsec:digraph-Maltsev-invariant}
	
	When the constraint language $\varGamma$ consists of a single binary relation $R\subseteq D^2$, the problem $\rcsp(\varGamma)$ coincides with the $H$-recoloring problem for the digraph $H=(D,R)$.
	The computational complexity of digraph recoloring has been extensively studied.
	The most general known result for graphs is due to \citet{Wro20}, who showed that $H$-recoloring is solvable in polynomial time when the graph $H$ is square-free, i.e., $H$ does not contain the 4-cycle as a subgraph.
	Building on this, \citet{LMS25} extended the result and techniques to digraphs, demonstrating that $H$-recoloring is also solvable in polynomial time when the digraph $H$ does not contain a certain digraph as a subgraph.
	In this subsection, we clarify how these results relate to binary relations that are invariant under ordered partial Maltsev operations.
	
	Here, we introduce several terminologies from graph theory.
	A \emph{directed graph (digraph)} is a pair $(V(G),A(G))$ where $V(G)$ is a finite set of vertices and $A(G) \subseteq V(G)\times V(G)$ are arcs. 
	Hence, the set of arcs can be regarded as a binary relation on $V(G)$.
	We write $u \rightarrow v$ when $(u,v) \in A(G)$. 
	We say that a digraph $G$ is \emph{symmetric} if $v \rightarrow u$ whenever $u \rightarrow v$. 
	We interpret a symmetric digraph as an undirected graph and think of two arcs $\{ u \rightarrow v,v \rightarrow u\}$ as an undirected edge, which we write as $uv$.
	We refer to an undirected graph simply as a graph.
	We write $E(G)$ for the set of undirected edges of a graph $G$.
	For any digraph $G$, we associate to $G$ a graph $\overline{G}$ where $V(\overline{G})=V(G)$ and $uv \in E(\overline{G})$ if $u\rightarrow v$ or $v\rightarrow u$. 
	A \emph{walk} $W$ in a graph $G$ is a sequence of consecutive edges $W = (v_1v_2)(v_2v_3)\dots(v_{n-1}v_n)$. 
	The length $|W|$ of $W$ is the number of edges of $W$. 
	A \emph{cycle} $C$ is a closed walk, i.e., a walk such that $v_1 =v_n$ on $n-1$ distinct vertices. 
	A walk in digraph $G$ is a walk in $\overline{G}$. 
	A walk $W =(v_1v_2)\dots(v_{n-1}v_n)$ in a digraph is \emph{directed} if $v_i \rightarrow v_{i+1}$ for all $1\leq i\leq n - 1$.
	
	We first introduce classes of digraphs characterized by (partial) Maltsev operations.
	
	We redefine the notion of rectangularity defined for binary relations in terms of digraphs, and then introduce a property called total rectangularity.
	\begin{definition}[{\citealp[Definition 8]{Kaz11}}]
    A digraph $G$ is \emph{rectangular} if $(u,x)\in A(G)$ whenever
	$(u,w),(v,w),(v,x)\in A(G)$.
	\end{definition}
	
	If $u$ and $v$ are vertices of a digraph $G$, $u \overset{k}{\to} v$ denotes the existence of a directed walk from $u$ to $v$ of length $k$.
	
	\begin{definition}[{\citealp[Definition 3.1]{CEJN15}}]
		A digraph G is \emph{$k$-rectangular} if, for all vertices $u,v,w,x$, $u \overset{k}{\to} w,  v \overset{k}{\to} w, v \overset{k}{\to} x$ imply $ u \overset{k}{\to} x$.
		A digraph is 
		\emph{totally rectangular} if it is $k$-rectangular for every $k \ge 1$.
	\end{definition}
	Note that a digraph is rectangular if and only if it is 1-rectangular.
	
	When the arc set of a digraph is regarded as a binary relation, and this relation is invariant under a certain operation, we say that the digraph is invariant under that operation.
	Digraphs invariant under a Maltsev operation is characterized in \citep{CEJN15}.
	
	\begin{theorem}[{\citealp[Corollary 4.11]{CEJN15}}]
		\label{thm:totally-rectangular<=>Maltsev}
		A digraph $G=(V(G),A(G))$ is totally rectangular if and only if it is invariant under some Maltsev operation on $V(G)$.
	\end{theorem}
	
	It is also observed in Observation 7 in \citep{Kaz11} that digraphs invariant under a Maltsev operation are rectangular.
	We strengthen this result and characterize the rectangular graphs by the partial Maltsev operation.
	
	\begin{theorem}
		\label{thm:rectangular<=>partial-Maltsev}
		A digraph $G=(V(G),A(G))$ is rectangular if and only if it is invariant under the partial Maltsev operation on $V(G)$.
	\end{theorem}
	\begin{proof}
		For the if part, assume that $G$ is invariant under the partial Maltsev operation on $V(G)$.
		Take any arcs $(u,w),(v,w),(v,x) \in A(G)$.
		Then applying the partial Maltsev operation $M_p$ yields $(u,x)$.
		As $G$ is invariant under $M_p$, we have $(u,x) \in A(G)$.
		
		For the only-if part, assume that $G$ is rectangular.
		Take any arcs $a_1=(x_1,y_1),a_2=(x_2,y_2),a_3=(x_3,y_3) \in A(G)$ such that the partial Maltsev operation $M_p$ is defined on those arcs.
		Then we have (i) $x_1 = x_2$ or $x_2 = x_3$, and (ii) $y_1 = y_2$ or $y_2 = y_3$.
		Assume that $x_2 = x_3$.
		Then we have $M_p(x_1,x_2,x_3) = x_1$.
		In the case of $y_1 = y_2$ we have $M_p(y_1,y_2,y_3) = y_3$, and 
		since $G$ is rectangular we have $(x_1,y_3) \in A(G)$, implying that $M_p(a_1,a_2,a_3) \in A(G)$.
		In the case of $y_2 = y_3$ we have $M_p(y_1,y_2,y_3) = y_1$, implying that $M_p(a_1,a_2,a_3) = (x_1,y_1) \in A(G)$.
		Similarly, we have $M_p(a_1,a_2,a_3) \in A(G)$ in the case of $x_1 = x_2$.
		Hence, $G$ is invariant under $M_p$.
	\end{proof}
	
	Next, we compare directed graphs for which the corresponding RCSP is known to be solvable in polynomial time with those that are invariant under an ordered partial Maltsev operation.
	First, we summarize the results from previous studies.
	
	As one of the most general results on $H$-recoloring for undirected graphs $H$, \citet{Wro20} established the following theorem.
	
	\begin{theorem}[{\citealp[Theorem 8.2]{Wro20}}]
		Let $H$ be a loopless graph that contains no 4-cycle as subgraph. 
		Then $H$-recoloring admits a polynomial-time algorithm.
	\end{theorem}
	
	We show that the 4-cycle, which is excluded in Wrochna’s result mentioned above, is invariant under a Maltsev operation.
	Before that we prepare an auxiliary lemma that is useful to show that graphs are $k$-rectangular for a positive integer $k$.
	For a vertex $x$ of a digraph and a positive integer $k$, let $x^{+k} = \{ y \mid  x \overset{k}{\to} y \}$.
	
	\begin{lemma}
		\label{lem:k-rectangularity-characterization}
		Let $k$ be an arbitrary positive integer.
		A digraph is $k$-rectangular if and only if for each vertex $u$ and $v$ either $u^{+k}=v^{+k}$ or $u^{+k} \cap v^{+k} = \emptyset$ holds.
	\end{lemma}
	
	\begin{proof}
		Assume that a digraph is $k$-rectangular.
		For two vertices $u,v$ of the graph, assume that $u^{+k} \cap v^{+k} \neq \emptyset$.
		Then for each $w \in v^{+k}$, by the definition of $k$-rectangularity, we have $w \in u^{+k}$. Thus we have $v^{+k} \subseteq u^{+k}$.
		Similarly, it can be verified that $u^{+k} \subseteq v^{+k}$.
		Hence, we have $u^{+k} = v^{+k}$.
		
		Conversely, assume that for each vertex $u$ and $v$ either $u^{+k}=v^{+k}$ or $u^{+k} \cap v^{+k} = \emptyset$ holds.
		Assume that we have $u \overset{k}{\to} w,  v \overset{k}{\to} w, v \overset{k}{\to} x$ for some vertices $u,v,w,x$.
		This means that $u^{+k} \cap v^{+k} \neq \emptyset$, and thus $u^{+k} = v^{+k}$.
		Therefore, $x \in u^{+k}$ holds, namely, we have $u \overset{k}{\to} x$.
	\end{proof}
	
	\begin{theorem}\label{thm:C_4-Maltsev-invariant}
		The 4-cycle is invariant under a Maltsev operation.
	\end{theorem}

	\begin{proof}
		By \cref{thm:totally-rectangular<=>Maltsev} it suffices to show that the 4-cycle is totally rectangular.
		It can be verified that for each pair of diagonally opposite vertices (say, $u$ and $v$) of the 4-cycle, 
		we have $u^{+k} = v^{+k}$.
		Moreover, for distinct vertices $u$ and $v$ that are not diagonally opposite, we have $x^{+k} \cap y^{+k} = \emptyset$.
		Therefore, the 4-cycle is $k$-rectangular by \cref{lem:k-rectangularity-characterization}. 
		Since $k$ is an arbitrary positive integer, the 4-cycle is totally rectangular.
	\end{proof}

	One of the consequences of \Cref{thm:ordered-Maltsev=>RCSP-is-P,thm:rectangular<=>partial-Maltsev} is that if a graph $H$ is rectangular, then $H$-recoloring is in P.
	Note that if a graph is rectangular, then $uw, vw, vx$ being edges implies that $ux$ is also an edge of the graph.
	It follows that whenever there exists a path $u-w-v-x$ of length three from vertex $u$ to $x$, there also exists an edge $u-x$.
	In this sense, a rectangular graph might contain many 
	4-cycles.

	Next, we examine the result by \citet{LMS25}, which extends Wrochna’s result to the digraph recoloring, and its connection to ordered partial Maltsev operations.
	
	An arc from a vertex to the same vertex (i.e., an arc of the form $(u,u)$) is called a \emph{loop}.
	A digraph $G$ is \emph{reflexive} if it has a loop on each vertex. 
	Given a digraph $G$, its \emph{reflexive closure}, denoted $r(G)$, is the reflexive graph obtained by adding a loop to every vertex lacking one.
	The \emph{algebraic girth} of a cycle $C$ in a digraph $G$ is the absolute value of the number of forward arcs minus the number of backward arcs.
	The two non-isomorphic orientations of a 4-cycle of algebraic girth zero, denoted by $\overrightarrow{C_4^1}$ and $\overrightarrow{C_4^2}$, are shown in \Cref{fig:cycle4}, 
	while the orientation of a 3-cycle of algebraic girth one, denoted by $\overrightarrow{K_3}$, is depicted in \Cref{fig:cycle3}.
	
	Generalizing Wrochna’s aforementioned result, \citet{LMS25} established the following two theorems, which represent the most general results on $H$-recoloring for digraphs $H$.
	
	\begin{figure}[t]
		\centering
		
		\begin{minipage}[t]{0.45\textwidth}
			\centering
			\begin{tikzpicture}
				\tikzset{
					vertex/.style={draw, circle, minimum size=14pt, inner sep=0pt},
					edge/.style={->, thick}
				}
				
				\node at (-1.25,2.75) {$\overrightarrow{C_4^1}$};
				\begin{scope}[rotate=45]
					\node[vertex] (A2) at (0,0) {$3$};
					\node[vertex] (B2) at (2,0) {$2$};
					\node[vertex] (C2) at (2,2) {$0$};
					\node[vertex] (D2) at (0,2) {$1$};
					\draw[edge] (B2) -- (A2);
					\draw[edge] (C2) -- (B2);
					\draw[edge] (C2) -- (D2);
					\draw[edge] (D2) -- (A2);
				\end{scope}
				
				\node at (2.25,2.75) {$\overrightarrow{C_4^2}$};
				\begin{scope}[shift={(3.5,0)}, rotate=45]
					\node[vertex] (A3) at (0,0) {3};
					\node[vertex] (B3) at (2,0) {2};
					\node[vertex] (C3) at (2,2) {0};
					\node[vertex] (D3) at (0,2) {1};
					\draw[edge] (A3) -- (B3);
					\draw[edge] (C3) -- (B3);
					\draw[edge] (C3) -- (D3);
					\draw[edge] (A3) -- (D3);
				\end{scope}
			\end{tikzpicture}
			\caption{The two non-isomorphic orientations of a 4-cycle of algebraic girth zero}
			\label{fig:cycle4}
		\end{minipage}
		\hfill
		\begin{minipage}[t]{0.45\textwidth}
			\centering
			\begin{tikzpicture}
				\tikzset{
					vertex/.style={draw, circle, minimum size=14pt, inner sep=0pt},
					edge/.style={->, thick}
				}
				
				\node at (0.25,2.75) {$\overrightarrow{K_3}$};
				\node[vertex] (A1) at (0,0) {0};
				\node[vertex] (B1) at (3,0) {2};
				\node[vertex] (C1) at (1.5,2.55) {1};
				\draw[edge] (A1) -- (B1);
				\draw[edge] (C1) -- (B1);
				\draw[edge] (A1) -- (C1);
			\end{tikzpicture}
			\caption{The orientation of a 3-cycle of algebraic girth one}
			\label{fig:cycle3}
		\end{minipage}
	\end{figure}
	
	\begin{theorem}[{\citealp[Theorem 1.1]{LMS25}}]
		\label{thm:LMS25-1}
		Let $H$ be a loopless digraph in which neither $\overrightarrow{C_4^1}$ nor $\overrightarrow{C_4^2}$ appears as subgraph. 
		Then $H$-Recoloring admits a polynomial-time algorithm.  
	\end{theorem}
	
	\begin{theorem}[{\citealp[Theorem 1.2]{LMS25}}]
		\label{thm:LMS25-2}
		Let $H$ be a reflexive digraph in which none of $\overrightarrow{K_3}$, $\overrightarrow{C_4^1}$ or $\overrightarrow{C_4^2}$ appears as subgraph.
		Then $H$-Recoloring admits a polynomial-time algorithm.
	\end{theorem}
	
	We show that $\overrightarrow{C_4^1}$ and $\overrightarrow{C_4^2}$ are invariant under a Maltsev operation.
	Later, we demonstrate that the reflexive closure $r(\overrightarrow{K_3})$ of $\overrightarrow{K_3}$ is invariant under an ordered partial Maltsev operation.
	
	\begin{proposition}
		\label{prop:C_4^1C_4^2-Maltsev-invariant}
		$\overrightarrow{C_4^1}$ and $\overrightarrow{C_4^2}$ are invariant under a Maltsev operation.
	\end{proposition}
	
	\begin{proof}
		It is straightforward to verify that these digraphs are totally rectangular. Therefore, by \cref{thm:totally-rectangular<=>Maltsev}, they are invariant under some Maltsev operation.
	\end{proof}
	
	Additionally, it has been established that $H$-recoloring is solvable in polynomial time when $H$ is a circular clique $C_{p,q}$ for $2 \le p/q <4$ \citep{BMMN16} and
	$H$ is a transitive tournament $\overrightarrow{K_n}$ for $n \ge 3$ \citep{DS23}. 
	Here, for integers $p>q >0$ with $p/q \ge 2$, the \emph{circular clique} $C_{p,q}$ is defined as the graph with vertex set $\{0,1,\dots, p-1\}$ and edge set $\{ ij \mid q \le |i-j| \le p-q \}$.
	Moreover, for an integer $n \ge 3$, the \emph{transitive tournament} $\overrightarrow{K_n}$ is defined as the digraph with vertex set $\{0,1,\dots, n-1\}$ and arc set $\{ (i,j) \mid i < j \}$.
	While these graphs do not meet the assumptions of \Cref{thm:LMS25-1,thm:LMS25-2}, \citet{LMS25} demonstrated that H-recoloring for such graphs can nevertheless be addressed through topological techniques.
	
	We show that among the circular cliques $C_{p,q}$ satisfying $2 \le p/q <4$, some are invariant under an ordered partial Maltsev operation while others are not.
	
	\begin{proposition}
		\label{prop:C_63-Maltsev}
		The circular clique $C_{6,3}$ is invariant under a Maltsev operation.
	\end{proposition}
	
	\begin{proof}
		The vertex set of $C_{6.3}$ is $\{ 0,1,\dots, 5 \}$ and the edge set of it is $\{ 03, 14, 25 \}$.
		It can be verified that for any distinct $u,v \in V(C_{6,3})$ and any positive integer $k$, we have $u^{+k} \cap v^{+k} = \emptyset$.
		Hence, by \cref{lem:k-rectangularity-characterization}, $C_{6.3}$ is totally rectangular, and thus invariant under a Maltsev operation by \cref{thm:totally-rectangular<=>Maltsev}.
	\end{proof}

	\begin{proposition}
		\label{prop:C_62-non-invariance}
		The circular clique $C_{6,2}$ is not invariant under any ordered partial Maltsev operation.
	\end{proposition}
	
	\begin{proof}
		Let $D=\{ 0,1,\dots,5\}$ and assume contrarily that there exists a total order $\le$ on $D$ such that $C_{6.2}$ is invariant under $M_{D,\le}$.
		Assume that $0 < 1$.
		As $04, 14, 15 \in E(C_{6.2})$ and $05 \notin E(C_{6.2})$, we have to have $4 < 5$, 
		since otherwise (i.e., if $5 < 4$), $C_{6.2}$ is not invariant under $M_{D,\le}$.
		Similarly, $03, 13, 15 \in E(C_{6.2})$ and $05 \notin E(C_{6.2})$ imply that $3 < 5$.
		Moreover, we have $41, 51, 53 \in E(C_{6.2})$ and $43 \notin E(C_{6.2})$, implying that $1 < 3$, and 
		$42, 52, 53 \in E(C_{6.2})$ and $43 \notin E(C_{6.2})$, implying that $2 < 3$.
		Finally, we have $25, 35, 31 \in E(C_{6.2})$ and $21 \notin E(C_{6.2})$, implying that $5 < 1$.
		Now, we reached to a contradiction that $1 < 3 < 5 < 1$.
		Similarly, in the case of $1 < 0$, we reach to a contradiction.
		Hence, $C_{6.2}$ is not invariant under $M_{D,\le}$.
	\end{proof}
	
	
	Finally, we show the following result on transitive tournaments.
	
	\begin{theorem}
		\label{thm:transitive-tournament-poMaltsev-invariant}
		For all $n \ge 3$, 
		the transitive tournament $\overrightarrow{K_n}$ and its reflexive closure $r(\overrightarrow{K_n})$ are invariant under an ordered partial Maltsev operation.
	\end{theorem}
	
	\begin{proof}
		We first show that the transitive tournament $\overrightarrow{K_n}$ is invariant under  $M_{[n],\le}$.
		Let $(i_t,j_t)\in A(\overrightarrow{K_n})$ for $t\in[3]$.
Assume that both $M_{[n],\le}(i_1,i_2,i_3)$ and $M_{[n],\le}(j_1,j_2,j_3)$ are defined.
		Without loss of generality, we assume that $i_1 \le i_2 = i_3$ and $j_1 = j_2 \ge j_3$.
		Then, 
		\begin{align}
			M_{[n],\le}\begin{pmatrix}
				i_1 & i_2 & i_3 \\
				j_1 & j_2 & j_3
			\end{pmatrix}
			= \begin{pmatrix}
				i_1 \\
				j_3 
			\end{pmatrix}.
		\end{align}
		Moreover, since $i_1 \le i_2$ and $i_2=i_3 < j_3$, we have $i_1 < j_3$.
		Hence, $(i_1,j_3) \in A(\overrightarrow{K_n})$.
		It follows that $\overrightarrow{K_n}$ is invariant under  $M_{[n],\le}$.
		
		Note that the arc set of the reflexive transitive tournament is given by $\{(i,j) \mid i \le j\}$.
		Consequently, analogous to the loopless case, we can show that the reflexive transitive tournament is invariant under $M_{[n],\le}$.
	\end{proof}

	\section{Safely componentwise bijunctive relations}
	\label{sec:SCB}
	
		In this section, we show that although $\varGamma_{\scb}$ can be characterized using partial operations, in contrast to the safely OR-free case, it cannot be characterized by a finite set of partial operations.
	
	We first show that $\varGamma_{\scb}$ can be characterized by partial polymorphisms.
	
	\begin{theorem}\label{lem:SCB-logical-characterization}
		We have $\varGamma_{\scb} = \langle \varGamma_{\scb} \rangle_{\wedge,=,\false}$.
		Thus, $\varGamma_{\scb} = \inv(\ppol(\varGamma_{\scb}))$. 
	\end{theorem}
	
	\ifconference
\else 
	\begin{proof}
		The inclusion $\varGamma_{\scb} \subseteq \langle \varGamma_{\scb} \rangle_{\wedge,=,\false}$ is trivial.
		Hence, we will show that $\varGamma_{\scb} \supseteq \langle \varGamma_{\scb} \rangle_{\wedge,=,\false}$.
		
		The equality relation $\Delta_{\{0,1\}}$ and the empty relations $\emptyset^{(1)}$ are safely componentwise bijunctive.
		Therefore, it suffices to show that 
		if two relations $R_1$ and $R_2$ are both safely componentwise bijunctive, then the relation expressed by $R_1(\bs{x}) \wedge R_2(\bs{x}')$ is also safely componentwise bijunctive, since once this is done, we can show any relation expressed by a $(\varGamma_{\scb} \cup \{\Delta_{\{0,1\}}\} \cup \{ \emptyset^{(1)} \})$-formula is safely componentwise bijunctive by induction on the number of relations in the formula.
		
		We will in fact show that if two relations $R_1$ (of arity $r_1$) and $R_2$ (of arity $r_2$) are both safely componentwise bijunctive, then the product $R_1 \times R_2$ is also safely componentwise bijunctive.
		Once this is done, we can conclude that 
		$R_1(\bs{x}) \wedge R_2(\bs{x}')$ 
		is also safely componentwise bijunctive, since it is obtained from $R_1 \times R_2$ by identification of variables (and a permutation of variables), which preserves safe componentwise bijunctivity.
		
		To show that $R=R_1 \times R_2$ is safely componentwise bijunctive, 
		we show that each relation $R'$ obtained from $R$ by identification of variables is componentwise bijunctive.
		
		Let $R'$ be a relation obtained from $R$ by identification of variables.
		Note that identification of variables can be realized by a sequence of identification of two variables.
		For example, identification of variable $(x_1,x_1,x_2,x_2,x_1)$ can be realized by a sequence $(x_1,x_1,x_3,x_4,x_5)$ (identification of $x_1$ and $x_2$), $(x_1,x_1,x_3,x_3,x_5)$ (identification of $x_3$ and $x_4$), $(x_1,x_1,x_3,x_3,x_1)$ (identification of $x_1$ and $x_5$) up to the renaming of variables.
		Moreover, we may identify two variables in any order, since the only thing that matters is which variable is finally identified with which variable.
		Hence, we may assume that the identification of variable for $R$ is first applied to variables appearing in $R_1$ (i.e., $x_1,\dots, x_{r_1}$), and then variables appearing in $R_2$ (i.e., $x_{r_1+1},\dots, x_{r_1+r_2}$), and finally variables contained in $R_1$ and variables contained in $R_2$ (i.e., identification of variables $x_i \in \{x_1,\dots, x_{r_1}\}$ and $x_j \in \{x_{r_1+1},\dots, x_{r_1+r_2}\}$), where we without loss of generality assume that these final identifications of variables do not induce any identification among the variables within $x_1,\dots, x_{r_1}$ or among those within $x_{r_1+1},\dots, x_{r_1+r_2}$.
		
		Let $R_1'$ (resp., $R_2'$) be the relation obtained from $R_1$ (resp., $R_2$) by the identification of variables appearing in $R_1$ (resp., $R_2$).
		Since $R_1$ and $R_2$ are safely componentwise bijunctive, $R_1'$ and $R_2'$ are componentwise bijunctive.
		
		Let $C \subseteq R'$ be a connected component of $G(R')$.
		We claim that there exist connected components $C_1$ and $C_2$ of $G(R_1')$ and $G(R_2')$, respectively, such that $C$ is contained in the relation obtained from $C_1 \times C_2$ by the identification of variables.
		Indeed, assume that $C_1'$, if it exists, be a connected component of $G(R_1')$ other than $C_1$.
		Let $\bs{s} \in C_1$ and $\bs{t} \in C'_1$.
		Then there exist two distinct indices $i$ and $j$ such that $s_i \neq t_i$ and $s_j \neq t_j$, since $C_1$ and $C'_1$ are disconnected in $G(R_1')$.
		Then $s_i \neq t_i$ and $s_j \neq t_j$ are preserved by the identification of variables in $R_1$ and variables in $R_2$ when constructing $R'$, since we assume that these identifications of variables do not induce any identification among the variables within $x_1,\dots, x_{r_1}$.
		It follows that $C_1$ and $C_1'$ remain disconnected after the identification of variables in $R_1$ and variables in $R_2$ when constructing $R'$.
		Similarly, two connected components $C_2$ and $C_2'$ remain disconnected after the identification of variables.
		Thus, $C$ is contained in the relation obtained from $C_1 \times C_2$ by the identification of variables.
		
		Now, since $C_1$ and $C_2$ are invariant and the majority operation on $\{0,1\}$ by \cref{lem:bijunctive<=>majority-invariant},  $C_1 \times C_2$ is also invariant under the majority operation.
		Moreover, every connected component of a relation invariant under the majority operation is also invariant under the majority operation by \cref{lem:connected_component_closedness}.
		Furthermore, identification of variables preserves invariance under an operation.
		Hence, $C$ is also invariant under the majority operation.
		This means that $C$ is bijunctive by \cref{lem:bijunctive<=>majority-invariant}.
		Therefore, $R$ is safely componentwise bijunctive.
		
		Now, the latter statement follows from \Cref{prop:pPol-logical-characterization}.
	\end{proof}
	\fi

	Unlike the case of the safely OR-free relations studied in the previous section, 
	we show that the safely componentwise bijunctive relations \emph{cannot} be characterized by any finite set of partial operations.
	
	\begin{theorem}\label{thm:SCB-not-finitely-many-pPol}
		Let $F$ be a finite set of partial operations.
		Then $\Gamma_\scb \neq \inv(F)$.
	\end{theorem}
	
	
	This theorem can be shown by using the following lemma.
	
	\begin{lemma}\label{lem:M^r-minimally-non-SCB}
		There exists an infinite family $(M^{(r)})_{r \ge 3}$ of relations 
		that are minimally not safely componentwise bijunctive, that is, $M^{(r)}$ is not safely componentwise bijunctive and for every relation $R \subsetneq M^{(r)}$ it holds that $R$ is safely componentwise bijunctive.
		Moreover, the cardinality of $M^{(r)}$ is $r+2$, i.e., $|M^{(r)}| = r+2$.
	\end{lemma}
	
	Using the above lemma, we can show \Cref{thm:SCB-not-finitely-many-pPol} as follows.
	
	\begin{proof}[of \Cref{thm:SCB-not-finitely-many-pPol}]
		Assume that there exists a finite set $F$ of partial operations such that $\Gamma_\scb = \inv(F)$.
		Let $k$ be the arity of an operation with the maximum arity in $F$.
		Consider a relation $M^{(r)}$ such that $|M^{(r)}| > k$.
		Since $M^{(r)}$ is not safely componentwise bijunctive, 
		there exist $f$ in $F$ such that $M^{(r)}$ is not invariant under $f$.
		Let $k' (\le k)$ be the arity of $f$.
		Then there exist $k'$ tuples in $M^{(r)}$ such that the tuple resulting from applying $f$ to these $k'$ tuples is defined and is not in $M^{(r)}$.
		However, since $M^{(r)}$ is minimally not safely componentwise bijunctive, and the relation $R$ consisting of these $k'$ tuples is a proper subset of $M^{(r)}$ (as $k' < r+2$), it holds that $R$ is safely componentwise bijunctive.
		This implies that $R$ is invariant under $f$ and that the tuple resulting from applying $f$ to these $k'$ tuples is in $R (\subseteq M^{(r)})$, a contradiction.
		Hence, $\Gamma_\scb \neq \inv(F)$.
	\end{proof}
	
	Hence, it remains to show \Cref{lem:M^r-minimally-non-SCB}.
	We use the following auxiliary lemma.
	For $r \ge 1$ and $o \in \{-,+\}^r$, 
	let us define a partial order $\le_{o}$ on $\{0,1\}^r$ as follows:
	for any two tuples $\bs{x},\bs{y}$ in $\{0,1\}^r$ we have $\bs{x} \le_o \bs{y}$ if and only if 
	$x_j \le y_j$ if $o_j=+$ and $x_j \ge y_j$ if $o_j=-$.
	We say that a relation $R \subseteq \{0,1\}^r$ \emph{admits a total order} if 
	there exists $o \in \{-,+\}^r$ such that 
	any two tuples $\bs{x},\bs{y}$ in $R$ are comparable under $\le_{o}$, that is, we have either $\bs{x} \le_o \bs{y}$ or $\bs{y} \le_o \bs{x}$.
	
	\begin{lemma}\label{lem:total-order=>SCB}
		Let $R$ be an $r$-ary relation on $\{0,1\}$.
		If $R$ admits a total order, then it is safely componentwise bijunctive.
	\end{lemma}
	\begin{proof}
		Let $o \in \{-,+\}^r$ be a tuple that witnesses that $R$ admits a total order.
		
		First, we show that $R$ is componentwise bijunctive.
		Let $C$ be a connected component of $R$.
		From \cref{lem:bijunctive<=>majority-invariant} in \Cref{subsec:results-for-Boolean-RCSP}, it suffices to show that for any $\bs{t}^1,\bs{t}^2,\bs{t}^3 \in C$ we have $\maj(\bs{t}^1,\bs{t}^2,\bs{t}^3) \in C$ for the unique majority operation $\maj$ on $\{0,1\}$.
		Take any tuples $\bs{t}^1,\bs{t}^2,\bs{t}^3 \in C$.
		By assumption, these tuples are comparable under $\le_o$.
		Assume without loss of generality that $\bs{t}^1 \le_o \bs{t}^2 \le_o \bs{t}^3$.
		We claim that $\maj(\bs{t}^1,\bs{t}^2,\bs{t}^3) = \bs{t}^2$.
		Indeed, for any index $j \in [r]$, $t^2_j$ is the median among the values $t^1_j$, $t^2_j$, and $t^3_j$.
		Thus, $\maj(t^1_j,t^2_j,t^3_j) = t^2_j$ for every $j \in [r]$ and $\maj(\bs{t}^1,\bs{t}^2,\bs{t}^3) = \bs{t}^2$ holds.
		Therefore, $\maj(\bs{t}^1,\bs{t}^2,\bs{t}^3) \in C$.
		
		Next, we show that any relation $R'$ obtained from $R$ by identification of variables is componentwise bijunctive.
		From the above argument, it suffices to show that $R'$ admits a total order.
		As we see in the proof of \cref{lem:SCB-logical-characterization}, 
		identification of variables can be realized by a sequence of identification of two variables.
		Hence, we may assume that $R'$ is obtained from $R$ by identification of two variables, say, $x_1$ and $x_r$ without loss of generality.
		Define $o' \in \{-,+\}^{r-1}$ as $o'_j=o_j$ for all $j \in [r-1]$.
		Now, take any tuples $\bs{t}^1,\bs{t}^2 \in R$ with $\bs{t}^1 \le_o \bs{t}^2$ and assume that these tuples remain after the identification of $x_1$ and $x_r$.
		This means that $t_1^1 = t_r^1$ and $t_1^2 = t_r^2$.
		Let $\bs{u}^1,\bs{u}^2$ be obtained from $\bs{t}^1,\bs{t}^2$, respectively, by the identification.
		Now, it holds that $\bs{u}^1 \le_{o'} \bs{u}^2$, 
		since $o$ and $o'$ agree in the coordinates in $[r-1]$ and $\bs{t}^1 \le_o \bs{t}^2$.
		Hence, any two tuples in $R'$ are comparable under $\le_{o'}$ and $R'$ admits a total order.
	\end{proof}
	
	Now we are ready to show \Cref{lem:M^r-minimally-non-SCB}.
	
	\begin{proof}[of \Cref{lem:M^r-minimally-non-SCB}]
		For each $r \ge 3$, we construct $M^{(r)}$ as 
		$M^{(r)} = \{ \bs{u}^{r,1}, \bs{u}^{r,2},\dots, \bs{u}^{r,r+2} \} \subseteq \{0,1\}^r$, 
		Here, $\bs{u}^{r,1} = (0,1,0,1,\dots,0,1)$ if $r$ is even and $\bs{u}^{r,1} = (0,1,0,1,\dots,0,1,0)$ if $r$ is odd.
		Namely, in the components of $\bs{u}^{r,1}$, 0 and 1 appear alternately.
		Now, for $2 \le i \le r+1$, $\bs{u}^{r,i}$ is defined recursively by 
		flipping the $(i-1)$th coordinate of $\bs{u}^{r,i-1}$.
		That is, $u^{r,i}_j=1-u^{r,i-1}_j$ if $j=i-1$, and $u^{r,i}_j=u^{r,i-1}_j$ if $j \neq i-1$.
		Finally, $\bs{u}^{r,r+2}$ is obtained from $\bs{u}^{r,r+1}$ by flipping its first coordinate, 
		that is, $u^{r,r+2}_j=1-u^{r,r+1}_j$ if $j=1$, and $u^{r,r+2}_j=u^{r,r+1}_j$ if $j \neq 1$.
		
		Examples of $M^{(r)}$ for small $r$'s are as follows:
		\begin{align*}
			&M^{(3)} = \{ (0,1,0), (1,1,0), (1,0,0), (1,0,1), (0,0,1) \},\\
			&M^{(4)} = \{ (0,1,0,1), (1,1,0,1), (1,0,0,1), (1,0,1,1), (1,0,1,0), (0,0,1,0) \},\\
			&M^{(5)} = \{ (0,1,0,1,0), (1,1,0,1,0), (1,0,0,1,0), (1,0,1,1,0), (1,0,1,0,0), (1,0,1,0,1), (0,0,1,0,1) \}.
		\end{align*}
		Note that $M^{(3)}$ is obtained from the relation $M$ defined in~\citep{GKMP09} by swapping the first and second coordinates.
		
		Now, we prove that $M^{(r)}$'s constructed as above are minimally not safely componentwise bijunctive.
		Fix $r \ge 3$.
		
		First, we show that $M^{(r)}$ is not safely componentwise bijunctive.
		It is easy to see that $G(M^{(r)})$ is connected as $\dist(\bs{u}^{r,i},\bs{u}^{r,i+1}) = 1$ for every $i \in [r+1]$ by construction.
		Moreover, we claim that $\bs{u}=\maj(\bs{u}^{r,1}, \bs{u}^{r,3}, \bs{u}^{r,r+2})$ is not in $M^{(r)}$.
		Indeed, $\bs{u}^{r,1}$ and $\bs{u}^{r,3}$ differ only in the first and second coordinates.
		Therefore, $u_j = u^{r,1}_j$ for all $j \ge 3$.
		Moreover, we have $u^{r,1}_1 = u^{r,r+2}_1 = 0$ and $u^{r,3}_2 = u^{r,r+2}_2 = 0$, implying that $(u_1,u_2) = (0,0)$.
		Since $(u^{r,i}_1,u^{r,i}_2) \neq (0,0)$ for all $i \in [r+1]$ and $u_3 = 0 \neq 1 = u^{r,r+2}_3$, 
		we conclude that $\bs{u}$ is not in $M^{(r)}$.
		Therefore, by \cref{lem:bijunctive<=>majority-invariant} in \Cref{subsec:results-for-Boolean-RCSP}, $M^{(r)}$ is not componentwise bijunctive and thus not safely componentwise bijunctive.
		
		Next, we show that for any $R \subsetneq M^{(r)}$ it holds that $R$ is safely componentwise bijunctive.
		Let $R \subsetneq M^{(r)}$.
		It is easy to see that each connected component of $R$ admits a total order.
		Moreover, let us assume that $R$ has two or more connected components.
		Assume that $\bs{u}^{r,i}$ and $\bs{u}^{r,k}$, where $1 \le i < k \le r+2$, are in different connected components, say $C_1$ and $C_2$.
		We will show that these tuples are never in the same connected component even after identification of variables is applied to $R$.
		Indeed, since $\bs{u}^{r,i}$ and $\bs{u}^{r,k}$ are disconnected, there exists $p \in \{i+1,\dots, k-1\}$ such that $\bs{u}^{r,p}$ is not in $R$.
		It follows that $k-i \ge 2$.
		Now, we show that there exist coordinates $j$ and $j'$ such that 
		$u^{r,i}_j = 0$ and $u^{r,k}_j = 1$, and $u^{r,i}_{j'} = 1$ and $u^{r,k}_{j'} = 0$.
		To show this, assume first that $k \neq r+2$.
		If $p$ is even, 
		then we see that $u^{r,i}_{p-1} = 0$ and $u^{r,k}_{p-1} = 1$, 
		and $u^{r,i}_{p} = 1$ and $u^{r,k}_{p} = 0$.
		If $p$ is odd, 
		then we see that $u^{r,i}_{p-1} = 1$ and $u^{r,k}_{p-1} = 0$, 
		and $u^{r,i}_{p} = 0$ and $u^{r,k}_{p} = 1$, as desired.
		The case of $k = r+2$ and $i\neq 1$ can be proven similarly.
		Finally, if $i=1$ and $k = r+2$, 
		then $u^{r,1}_{2} = 1$ and $u^{r,r+2}_{2} = 0$, 
		and $u^{r,1}_{3} = 0$ and $u^{r,r+2}_{3} = 1$, as desired.
		Now, we show that $\bs{u}^{r,i}$ and $\bs{u}^{r,k}$ are never in the same connected component even after identification of variables is applied to $R$.
		This is because, even if identification of variables is applied to $\bs{u}^{r,i}$ and $\bs{u}^{r,k}$, 
		$u^{r,i}_j = 0$ and $u^{r,k}_j = 1$, and $u^{r,i}_{j'} = 1$ and $u^{r,k}_{j'} = 0$ are preserved.
		Hence, the distances between the tuples in $C_1$ and $C_2$ are at least two and they remain disconnected.
		Hence, after the application of variable identification to $R$, 
		each connected component still admits a total order as shown in the proof of \cref{lem:total-order=>SCB}.
		Therefore, it is also componentwise bijunctive.
	\end{proof}

	
	\section{Conclusion}
	\label{sec:conclusion}
	We have presented evidence that the algebraic approach holds promise for analyzing the computational complexity of the reconfiguration CSP (RCSP). However, the theory of partial operations remains less developed than that of total operations. In the context of CSPs, pp-interpretations---generalizations of pp-definitions---are characterized by equalities satisfied by operations \citep[see, e.g.,][]{BKW17}. Whether similar characterizations extend to partial operations is still unclear and crucial for validating the algebraic framework.

	Recent CSP research has begun to explore the topological structure of solution spaces and its connection to complexity~\citep{SW24,Mey24,MO25}. 
	Notably, Wrochna’s graph recoloring algorithm~\citep{Wro20} and subsequent analyses~\citep{LMS25,FIK+25,FIK26,Mat26} 
	highlight the relevance of topological methods. 
	A promising direction for RCSP complexity analysis may lie in combining algebraic and topological approaches.
	
	\acknowledgements
	We thank Soichiro Fujii, Yuni Iwamasa, Yuta Nozaki, and Akira Suzuki for many insightful discussions. In particular, the proof of \Cref{lem:SCB-logical-characterization} benefited greatly from discussions with Akira Suzuki, and the proof of \Cref{thm:SCB-not-finitely-many-pPol} from discussions with Soichiro Fujii. We are also grateful to the anonymous reviewers for their constructive and valuable comments.

    \bibliographystyle{abbrvnat}
	\bibliography{reconfiguration}

@Preamble{ {\providecommand{\noopsort}[1]{}} }

@BOOK{Dec03,
  title = {Constraint Processing},
  publisher = {Morgan Kaufmann},
  year = {2003},
  author = {Rina Dechter},
  address = {California, USA},
  doi = {10.1016/B978-1-55860-890-0.X5000-2}
}

@Article{GKMP09,
  author    = {Parikshit Gopalan and Phokion G. Kolaitis and Elitza Maneva and Christos H. Papadimitriou},
  title     = {The Connectivity of {B}oolean Satisfiability: Computational and Structural Dichotomies},
  journal   = {SIAM Journal on Computing},
  year      = {2009},
  volume    = {38},
  pages     = {2330-2355},
  doi = {10.1137/07070440X}
}

@ARTICLE{Kaz11,
  author = {Alexandr Kazda},
  title = {Maltsev digraphs have a majority polymorphism},
  journal = {European Journal of Combinatorics},
  year = {2011},
  volume = {32},
  pages = {390-397},
  doi = {10.1016/j.ejc.2010.11.002}
}

@INPROCEEDINGS{Sch78,
  author = {Thomas J. Schaefer},
  title = {The Complexity of Satisfiability Problems},
  booktitle = {Proceedings of the 10th annual ACM Symposium on Theory of Computing},
  year = {1978},
  pages = {216-226},
  bibsource = {DBLP, http://dblp.uni-trier.de},
  comment = {DBLP:conf/stoc/STOC10},
  doi = {10.1145/800133.804350}
}

@Article{JLNZ17,
  author  = {Peter Jonsson and Victor Lagerkvist and Gustav Nordh and Bruno Zanuttini},
  title   = {Strong partial clones and the time complexity of {SAT} problems},
  journal = {Journal of Computer and System Sciences},
  year    = {2017},
  volume  = {84},
  pages   = {52-78},
  doi = {10.1016/j.jcss.2016.07.008}
}

@Book{CH99,
  author    = {Vijay Chandru and John Hooker},
  title     = {Optimization Methods for Logical Inference},
  year      = {1999},
  publisher = {Wiley-Interscience, New York, USA},
  doi = {10.1002/9781118033166}
}

@InProceedings{Zhu17,
  author    = {Dmitriy Zhuk},
  title     = {A Proof of {CSP} Dichotomy Conjecture},
  booktitle = {Proceedings of the 58th IEEE Annual Symposium on Foundations of Computer Science},
  year      = {2017},
  pages     = {331-342},
  doi = {10.1109/FOCS.2017.38}
}

@InProceedings{Bul17,
  author    = {Andrei A. Bulatov},
  title     = {A Dichotomy Theorem for Nonuniform {CSP}s},
  booktitle = {Proceedings of the 58th IEEE Annual Symposium on Foundations of Computer Science},
  year      = {2017},
  pages     = {319-330},
  doi={10.1109/FOCS.2017.37}
}

@article{IDHPSUU11,
	author = "Takehiro Ito and Erik D. Demaine and Nicholas J.A. Harvey and Christos H. Papadimitriou and Martha Sideri and Ryuhei Uehara and Yushi Uno",
	title = "On the complexity of reconfiguration problems",
	journal = "Theoretical Computer Science",
	volume = "412",
	number = "12--14",
	pages = "1054--1065",
	year = "2011",
    doi = {10.1016/j.tcs.2010.12.005}
    }

@Article{Sch14,
  author  = {Konrad W. Schwerdtfeger},
  title   = {A Computational Trichotomy for Connectivity of {B}oolean Satisfiability},
  journal = {Journal on Satisfiability, Boolean Modeling and Computation},
  year    = {2014},
  volume  = {8},
  pages   = {173-195},
  doi = {10.3233/SAT190097}
}

@Article{KiS21,
  author  = {Kei Kimura and Akira Suzuki},
  title   = {Trichotomy for the reconfiguration problem of integer linear systems},
  pages   = {88-109},
  volume  = {856},
  journal = {Theoretical Computer Science},
  year    = {2021},
  doi = {10.1016/j.tcs.2020.12.025}
}

@InCollection{BKW17,
  author    = {Libor Barto and Andrei Krokhin and Ross Willard},
  booktitle = {The Constraint Satisfaction Problem: Complexity and Approximability},
  title     = {Polymorphisms, and how to use them},
  editor    = {Andrei Krokhin and Stanislav \v{Z}ivn\'{y}},
  maintitle = {Dagstuhl Follow-Ups},
  pages     = {45-77},
  publisher = {Schloss Dagstuhl-Leibniz-Zentrum fuer Informatik},
  volume    = {7},
  year      = {2017},
  doi= {10.4230/DFU.Vol7.15301.1}
}

@ARTICLE{CEJN15,
  author = {Catarina Carvalho and L\'{a}szl\'{o} Egri and  Marcel Jackson and Todd Niven},
  title = {On Maltsev digraphs},
  journal = {The Electric Journal of Combinatorics},
  year = {2015},
  volume = {22},
  pages = {1-32},
  doi = {10.37236/4419}
}

@InProceedings{DR10,
  author    = {Martin E. Dyer and David M. Richerby},
  booktitle = {Proceedings of the forty-second ACM symposium on Theory of computing},
  year      = {2010},
  title     = {On the complexity of $\#${CSP}},
  pages     = {725-734},
  doi = {10.1145/1806689.1806789}
}

@ARTICLE{DS23,
  author = {Anton Dochtermann and Anurag Singh},
  title = {Homomorphism complexes, reconfiguration, and homotopy for directed graphs},
  journal = {European Journal of Combinatorics},
  year = {2023},
  volume = {110},
  pages = {1-31},
  doi = {10.1016/j.ejc.2023.103704}
}

@ARTICLE{Wro20,
  author = {Marcin Wrochna},
  title = {Homomorphism Reconfiguration via Homotopy},
  journal = {SIAM Journal on Discrete Mathematics},
  year = {2020},
  volume = {34},
  pages = {328-350},
  doi = {10.1137/17M1122578}
}

@ARTICLE{LW21,
  author = {Victor Lagerkvist and Magnus Wahlstr\"{o}m},
  title = {The (Coarse) Fine-Grained Structure of {NP}-Hard {SAT} and {CSP} Problems},
  journal = {ACM Transactions on Computation Theory},
  year = {2021},
  volume = {14},
  pages = {1-54},
  doi = {10.1145/3492336}
}

@article{LMS25,
  title={Reconfiguration of digraph homomorphisms},
  author={L{\'e}v{\^e}que, Benjamin and M{\"u}hlenthaler, Moritz and Suzan, Thomas},
  journal={SIAM Journal on Discrete Mathematics},
  volume={39},
  number={1},
  pages={327--360},
  year={2025},
  publisher={SIAM},
doi={10.1137/23M1623690}
}

@article{CvJ11,
  title={Finding paths between 3-colorings},
  author={Cereceda, Luis and van den Heuvel, Jan and Johnson, Matthew},
  journal={Journal of graph theory},
  volume={67},
  number={1},
  pages={69--82},
  year={2011},
  publisher={Wiley Online Library},
doi={10.1002/jgt.20514}
}

@article{BBD+21,
  title={Dismantlability, connectedness, and mixing in relational structures},
  author={Brice{\~n}o, Raimundo and Bulatov, Andrei and Dalmau, V{\'\i}ctor and Larose, Beno{\^\i}t},
  journal={Journal of Combinatorial Theory, Series B},
  volume={147},
  pages={37--70},
  year={2021},
  publisher={Elsevier},
doi={10.1016/j.jctb.2020.10.001}
}

@article{HIZ18,
  title={Complexity of reconfiguration problems for constraint satisfaction},
  author={Hatanaka, Tatsuhiko and Ito, Takehiro and Zhou, Xiao},
  journal={arXiv preprint arXiv:1812.10629},
  year={2018},
  doi = {10.48550/arXiv.1812.10629}
}

@article{Rom81,
  title={The algebras of partial functions and their invariants},
  author={Romov, Boris A},
  journal={Cybernetics},
  volume={17},
  number={2},
  pages={157--167},
  year={1981},
  publisher={Springer},
doi={10.1007/BF01069627}
}

@article{Gei68,
  title={Closed systems of functions and predicates},
  author={Geiger, David},
  journal={Pacific journal of mathematics},
  volume={27},
  number={1},
  pages={95--100},
  year={1968},
  publisher={Mathematical Sciences Publishers},
doi={10.2140/pjm.1968.27.95}
}

@inproceedings{SW24,
  title={A Topological Version of Schaefer’s Dichotomy Theorem},
  author={Schnider, Patrick and Weber, Simon},
  booktitle={40th International Symposium on Computational Geometry (SoCG 2024)},
  pages={77--1},
  year={2024},
  organization={Schloss Dagstuhl--Leibniz-Zentrum f{\"u}r Informatik},
doi={10.4230/LIPIcs.SoCG.2024.77}
}

@article{Mey24,
  title={A Dichotomy for Finite Abstract Simplicial Complexes},
  author={Meyer, Sebastian},
  journal={arXiv preprint arXiv:2408.08199},
  year={2024},
  doi = {10.48550/arXiv.2408.08199}
}

@inproceedings{MO25,
  title={A topological proof of the Hell-Ne{\v{s}}et{\v{r}}il dichotomy},
  author={Meyer, Sebastian and Opr{\v{s}}al, Jakub},
  booktitle={Proceedings of the 2025 Annual ACM-SIAM Symposium on Discrete Algorithms (SODA)},
  pages={4507--4519},
  year={2025},
  organization={SIAM},
doi={10.1137/1.9781611978322.154}
}

@inproceedings{JCG95,
  title={A unifying framework for tractable constraints},
  author={Jeavons, Peter and Cohen, David and Gyssens, Marc},
  booktitle={Principles and Practice of Constraint Programming—CP'95: First International Conference, CP'95 Cassis, France, September 19--22, 1995 Proceedings 1},
  pages={276--291},
  year={1995},
  organization={Springer},
doi={10.1007/3-540-60299-2_17}
}

@article{BMMN16,
  title={A dichotomy theorem for circular colouring reconfiguration},
  author={Brewster, Richard C and McGuinness, Sean and Moore, Benjamin and Noel, Jonathan A},
  journal={Theoretical Computer Science},
  volume={639},
  pages={1--13},
  year={2016},
  publisher={Elsevier},
doi={10.1016/j.tcs.2016.05.015}
}

@article{Jea98,
  title={On the algebraic structure of combinatorial problems},
  author={Jeavons, Peter},
  journal={Theoretical Computer Science},
  volume={200},
  number={1-2},
  pages={185--204},
  year={1998},
  publisher={Elsevier},
doi={10.1016/S0304-3975(97)00230-2}
}

@article{Zhu20,
  title={A proof of the {CSP} dichotomy conjecture},
  author={Zhuk, Dmitriy},
  journal={Journal of the ACM (JACM)},
  volume={67},
  number={5},
  pages={1--78},
  year={2020},
  publisher={ACM New York, NY, USA},
doi={10.1145/3402029}
}

@article{FIK+25,
  title={Homotopy types of {H}om complexes of graph homomorphisms whose codomains are cycles},
  author={Fujii, Soichiro and Iwamasa, Yuni and Kimura, Kei and Nozaki, Yuta and Suzuki, Akira},
  journal={Journal of Applied and Computational Topology},
  volume={9},
  number={21},
  year={2025},
doi={10.1007/s41468-025-00219-7}
}

@article{FIK26,
title={Homotopy types of {H}om complexes of graph homomorphisms whose codomains are square-free},
author={Fujii, Soichiro and Kimura, Kei and Nozaki, Yuta},
journal = {European Journal of Combinatorics},
volume = {131},
pages = {104238},
year = {2026},
doi = {10.1016/j.ejc.2025.104238},
}

@article{Mat26,
title = {Hom complexes of graphs whose codomains are square-free},
journal = {Journal of Combinatorial Theory, Series B},
volume = {178},
pages = {267-293},
year = {2026},
issn = {0095-8956},
doi = {10.1016/j.jctb.2026.01.005},
author = {Takahiro Matsushita},
}

@article{BC09,
  title={Finding paths between graph colourings: {PSPACE}-completeness and superpolynomial distances},
  author={Bonsma, Paul and Cereceda, Luis},
  journal={Theoretical Computer Science},
  volume={410},
  number={50},
  pages={5215--5226},
  year={2009},
  publisher={Elsevier},
  doi={10.1016/j.tcs.2009.08.023}
}

@article{LNS23,
  title={Recolouring homomorphisms to triangle-free reflexive graphs},
  author={Lee, Jae baek and Noel, Jonathan A and Siggers, Mark},
  journal={Journal of Algebraic Combinatorics},
  volume={57},
  number={1},
  pages={53--73},
  year={2023},
  publisher={Springer},
doi={10.1007/s10801-022-01161-y}
}

@article{BLS18,
  title={Recolouring reflexive digraphs},
  author={Brewster, Richard C and Lee, Jae-Baek and Siggers, Mark},
  journal={Discrete Mathematics},
  volume={341},
  number={6},
  pages={1708--1721},
  year={2018},
  publisher={Elsevier},
doi={10.1016/j.disc.2018.03.006}
}

@article{LNS20,
  title={Reconfiguring graph homomorphisms on the sphere},
  author={Lee, Jae-Baek and Noel, Jonathan A and Siggers, Mark},
  journal={European Journal of Combinatorics},
  volume={86},
  pages={103086},
  year={2020},
  publisher={Elsevier}, 
doi={10.1016/j.ejc.2020.103086}
}

@book{Ziv12,
  title={The complexity of valued constraint satisfaction problems},
  author={Zivny, Stanislav},
  year={2012},
  publisher={Springer Science \& Business Media},
  doi = {10.1007/978-3-642-33974-5}
}

@article{LW20,
  title={Sparsification of {SAT} and {CSP} problems via tractable extensions},
  author={Lagerkvist, Victor and Wahlstr{\"o}m, Magnus},
  journal={ACM Transactions on Computation Theory (TOCT)},
  volume={12},
  number={2},
  pages={1--29},
  year={2020},
  doi= {10.1145/3389411}
}

@article{Marx2005,
  author  = {Marx, D{\'a}niel},
  title   = {Parameterized complexity of constraint satisfaction problems},
  journal = {Computational Complexity},
  volume  = {14},
  pages   = {153--183},
  year    = {2005},
  publisher = {Springer},
  doi = {10.1007/s00037-005-0195-9}
}

@InProceedings{JLdW26,
  author =	{Jonsson, Peter and Lagerkvist, Victor and de Vlas, Jorke M. and Wahlstr\"{o}m, Magnus},
  title =	{{Going Beyond Twin-Width? CSPs with Unbounded Domain and Few Variables}},
  booktitle =	{53rd International Colloquium on Automata, Languages, and Programming (ICALP 2026)},
  pages =	{120:1--120:20},
  series =	{Leibniz International Proceedings in Informatics (LIPIcs)},
  year =	{2026},
  volume =	{374},
  doi =		{10.4230/LIPIcs.ICALP.2026.120},
}

@article{KSTW01,
author = {Khanna, Sanjeev and Sudan, Madhu and Trevisan, Luca and Williamson, David P.},
title = {The Approximability of Constraint Satisfaction Problems},
journal = {SIAM Journal on Computing},
volume = {30},
number = {6},
pages = {1863-1920},
year = {2001},
doi = {10.1137/S0097539799349948},
}

@inproceedings{Raghavendra08,
author = {Raghavendra, Prasad},
title = {Optimal algorithms and inapproximability results for every CSP?},
year = {2008},
doi = {10.1145/1374376.1374414},
booktitle = {Proceedings of the Fortieth Annual ACM Symposium on Theory of Computing},
pages = {245–254},
numpages = {10},
series = {STOC '08}
}

@article{BG21,
author = {Brakensiek, Joshua and Guruswami, Venkatesan},
title = {Promise Constraint Satisfaction: Algebraic Structure and a Symmetric Boolean Dichotomy},
journal = {SIAM Journal on Computing},
volume = {50},
number = {6},
pages = {1663-1700},
year = {2021},
doi = {10.1137/19M128212X},
}

@article{BBKO21,
author = {Barto, Libor and Bul\'{\i}n, Jakub and Krokhin, Andrei and Opr\v{s}al, Jakub},
title = {Algebraic Approach to Promise Constraint Satisfaction},
year = {2021},
issue_date = {August 2021},
volume = {68},
number = {4},
doi = {10.1145/3457606},
journal = {J. ACM},
numpages = {66},
}
	
\end{document}